\documentclass[10pt,letterpaper]{article}
\usepackage[top=0.85in,left=1in,right=1in,footskip=0.75in]{geometry}

\usepackage{amsmath,amssymb}

\usepackage{changepage}

\usepackage{textcomp,marvosym}
\usepackage{siunitx} 

\usepackage{cite}

\usepackage{nameref,hyperref}

\usepackage[right]{lineno}

\usepackage[nopatch=eqnum]{microtype}
\usepackage{iftex}
\ifPDFTeX
\DisableLigatures[f]{encoding = *, family = * }
\fi

\usepackage[table]{xcolor}

\usepackage{array}

\newcolumntype{+}{!{\vrule width 2pt}}

\newlength\savedwidth
\newcommand\thickcline[1]{%
  \noalign{\global\savedwidth\arrayrulewidth\global\arrayrulewidth 2pt}%
  \cline{#1}%
  \noalign{\vskip\arrayrulewidth}%
  \noalign{\global\arrayrulewidth\savedwidth}%
}

\newcommand\thickhline{\noalign{\global\savedwidth\arrayrulewidth\global\arrayrulewidth 2pt}%
\hline
\noalign{\global\arrayrulewidth\savedwidth}}

\raggedright
\usepackage[aboveskip=1pt,labelfont=bf,labelsep=period,justification=raggedright,singlelinecheck=off]{caption}
\renewcommand{\figurename}{Fig}

\makeatletter
\renewcommand{\@biblabel}[1]{\quad#1.}
\makeatother

\usepackage{lastpage,fancyhdr,graphicx}
\usepackage{epstopdf}
\fancyheadoffset[L]{0in}
\fancyfootoffset[L]{0in}
\DeclareSIUnit{\dyne}{dyn}
\DeclareSIUnit{\mmHg}{mmHg} 

\begin{document}
\vspace*{0.2in}

\begin{flushleft}
{\Large
\textbf\newline{svMultiPhysics: a finite element--based solver for cardiovascular simulations} 
}
\newline
\\
Sujal Dave\textsuperscript{1 \Yinyang},
David Codoni\textsuperscript{2 \Yinyang},
Javiera Jilberto\textsuperscript{1},
Michele Bucelli\textsuperscript{3},
Aaron L. Brown\textsuperscript{4,5},
Divya Adil\textsuperscript{4},
Zachary A. Sexton\textsuperscript{5,6},
Bahetihazi Maidu\textsuperscript{1},
Han Zhao\textsuperscript{1},
David W. Parker\textsuperscript{7},
Nathalie D'Amours\textsuperscript{1},
Martin R. Pfaller\textsuperscript{8},
Shawn C. Shadden\textsuperscript{9},
Charles A. Taylor\textsuperscript{3},
Alison L. Marsden\textsuperscript{1,4,5,6 *}

\bigskip
\textbf{1} Department of Pediatrics (Cardiology), Stanford University, Stanford, CA, USA
\\
\textbf{2} Energy Systems and Components Optimization, Water \& Energy Transition Unit, Flemish Institute for Technological Research (VITO), Boeretang 200, Mol, 2400, Belgium
\\
\textbf{3} Oden Institute for Computational Engineering and Sciences, University of Texas, Austin, TX, USA
\\
\textbf{4} Department of Mechanical Engineering, Stanford University, Stanford, CA, USA
\\
\textbf{5} Stanford Cardiovascular Institute, Stanford, CA, USA
\\
\textbf{6} Department of Bioengineering, Stanford University, Stanford, CA, USA
\\
\textbf{7} Stanford Research Computing Center, Stanford University, Stanford, CA, USA
\\
\textbf{8} Department of Biomedical Engineering, Yale University, New Haven, CT, USA
\\
\textbf{9} Department of Mechanical Engineering, University of California, Berkeley, CA, USA
\\
\bigskip

%
%
 \Yinyang These authors contributed equally to this work.





* amarsden@stanford.edu

\end{flushleft}
\section*{Abstract}
Heart disease remains the leading cause of death in the United States, motivating extensive efforts to improve its diagnosis, treatment, and prevention. Over the past decade, computational modeling has emerged as a powerful tool to advance cardiovascular research by enabling detailed, patient-specific studies of cardiac physiology and pathology.
svMultiPhysics is an open-source, parallel finite element solver written in C++ specifically designed for multiphysics cardiovascular problems. It provides a unified framework for simulating the partial differential equations that govern solid mechanics, fluid dynamics, diffusion, and cardiac electrophysiology. These equations can be solved independently or in a coupled fashion, allowing researchers to investigate interactions between physical processes in a modular yet integrated way.

We verify the implementation against published benchmarks for incompressible flow, fluid--structure interaction, cardiac electrophysiology, and cardiac mechanics, including comparisons with published results of its Fortran predecessor. A whole-heart fluid--structure interaction (FSI) example demonstrates prescribed active myocardial contraction coupled to blood flow, immersed valve surfaces, and a closed-loop circulation. Performance studies compare linear solvers and preconditioners for rigid-wall computational fluid dynamics (CFD) and FSI simulations of an abdominal aortic aneurysm (AAA) and biventricular electrophysiology simulations. Preliminary CPU--GPU comparisons show reduced runtimes for selected GPU-enabled configurations, subject to the limitations of short timing windows and different hardware architectures. As the primary three-dimensional solver in the SimVascular project, svMultiPhysics supports image-based cardiovascular modeling and is openly developed on GitHub.

\section*{Author summary}
In this work, we present an overview of svMultiPhysics, an open-source C++ parallel finite element solver for cardiovascular simulation and a core component of the SimVascular project for patient-specific modeling. The solver provides a unified framework for fluid dynamics, solid mechanics, fluid--structure interaction, and cardiac electrophysiology, and couples them within a consistent numerical infrastructure. Our goal is to describe the core features and modeling capabilities of the solver in an accessible way, complementing the available web-based documentation. We verify the implementation against published benchmark problems for incompressible flow, fluid--structure interaction, cardiac electrophysiology, and cardiac mechanics, including comparisons with results from its Fortran predecessor. We also demonstrate a whole-heart fluid--structure interaction simulation in which prescribed active myocardial contraction drives blood flow through immersed valve surfaces into a closed-loop circulation, and an idealized left-ventricle electromechanics example in which contraction is driven by the computed calcium transient and coupled to a lumped-parameter circulation. We also report the computational performance of three realistic examples: a computational fluid dynamics simulation of an abdominal aortic aneurysm, the corresponding fluid--structure interaction simulation, and a biventricular electrophysiology simulation. The availability of model and benchmark inputs is described in the Data availability section. By combining an overview of the methodology with verified and reproducible examples, we aim to give researchers a clear understanding of how svMultiPhysics can support advanced cardiovascular modeling.

\section*{Introduction}
Computational modeling has become an essential tool in cardiovascular research and clinical decision-making. Advances in imaging and diagnostic technologies, such as echocardiography, cardiac computed tomography (CT), magnetic resonance imaging (MRI), and electrocardiogram (ECG), now provide increasingly rich data describing complementary aspects of cardiovascular anatomy and function~\cite{sun2025}. However, each modality captures only a partial view of the underlying physics and physiology, and imaging alone cannot predict how a patient may respond to a proposed intervention. Physics-based numerical models can bridge these gaps by integrating clinical data with the governing equations of cardiovascular mechanics, electrophysiology, and hemodynamics. This integration enables both the simulation of quantities that are difficult or impossible to measure directly, including transmembrane potentials, tissue deformation, and detailed blood-flow patterns, and the prediction of how these quantities may change under alternative disease states, device designs, or treatment strategies~\cite{vardhan2021,schwarz2023}. Such models support clinicians in understanding disease mechanisms, anticipating hemodynamic responses to interventions, and evaluating treatment strategies in a controlled and non-invasive environment.

A wide spectrum of open-source computational frameworks have been used for cardiovascular or biomechanical and blood flow modeling. Examples include the well-established OpenFOAM~\cite{openfoam2009} solver for general computational fluid dynamics and FEBio~\cite{febio2012} for solid and soft-tissue biomechanics. They are often built on general finite-element libraries such as FEniCS~\cite{fenics2014}, and deal.II~\cite{dealii2021,dealii2025}. Several specialized cardiac and vascular modeling environments also exist: Chaste~\cite{chaste2009} offers a rigorously validated biology--oriented infrastructure; lifex~\cite{lifex2022,bucelli2025} extends deal.II to support detailed cardiac electrophysiology~\cite{lifexep2023}, mechanics~\cite{fedele2023}, and hemodynamics~\cite{lifexcfd}; CHeart~\cite{cheart} and 4C~\cite{4c} provide comprehensive multiphysics capabilities for cardiovascular problems; openCARP~\cite{opencarp} is an open cardiac electrophysiology simulator for in-silico experiments; TorchCor~\cite{zhou2026} provides high-performance GPU-accelerated cardiac electrophysiology simulations using the finite element method; OpenCOR~\cite{opencor2015} provides a modular environment for CellML-based computational biology models, including cardiac cell-scale models; the FEniCS-based OASIS solver~\cite{kjeldsberg2023} enables verified and validated moving-domain cardiovascular CFD simulations; CRIMSON~\cite{crimson} is an advanced simulation environment for subject-specific hemodynamic analysis; and IBAMR~\cite{griffith2017ibamr} provides a widely used, scalable open-source implementation of the immersed boundary method for FSI, and has recently been used for whole-heart FSI simulations with fully resolved valves and active myocardial contraction~\cite{davey2024wholeheart,sadipour2026tmvr}. Beyond open-source tools, Alya, developed at the Barcelona Supercomputing Center, provides comprehensive high-performance multiphysics capabilities and has been applied to large-scale cardiac electromechanics and fully coupled fluid--electro-mechanical heart simulations~\cite{vazquez2016alya,santiago2018alya}. These frameworks demonstrate the maturity of open-source cardiovascular modeling. Nevertheless, some tools are primarily designed around few dominant physics, often requiring additional software layers or user-implemented coupling strategies to perform multiphysics simulations. Cardiovascular problems are inherently multidisciplinary applications and this requires the development of frameworks designed with multiphysics integration in mind, where different processes share consistent discretizations, data structures, and solver strategies, with minimal user intervention.

In this work, we introduce svMultiPhysics, an open-source multiphysics parallel-distributed finite element solver implemented in C++. It provides a unified framework in which fluid dynamics, solid mechanics, scalar transport, and electrophysiology can be simulated within a consistent variational finite element formulation and naturally coupled when required. The solver integrates advanced linear algebra libraries such as PETSc~\cite{petsc} and Trilinos~\cite{trilinos}, enabling high-performance simulations across a range of computing architectures. The code base is in active development with current efforts focused on modularity, extensibility, and improved support for coupled physics. The solver is part of the SimVascular project~\cite{simvascular}, which provides a complete pipeline from medical image segmentation to patient-specific cardiovascular simulation and analysis.

In addition to CPU-based parallel execution, recent development efforts have enabled svMultiPhysics to be built with GPU-accelerated linear algebra backends through Kokkos programming modules in Trilinos~\cite{trott2021kokkos}. When compiled with appropriate device backends (e.g. CUDA), the solver leverages GPU execution for Krylov-based solvers and preconditioners operating on large linear systems~\cite{mayr2025trilinos}. This capability allows svMultiPhysics to take advantage of hybrid high performance computing (HPC) architectures. The preliminary CPU--GPU comparison for the abdominal aorta aneurysm (AAA) CFD case is described in the \nameref{sec:aaa-gpu-performance} subsection.

The svMultiPhysics solver is implemented in C++, a popular general-purpose object-oriented programming language that
provides powerful features for producing modular software with a clear structure. It was developed as a new implementation
of the svFSI~\cite{svfsi} multiphysics solver written in Fortran 90, released previously as part of SimVascular.  Although Fortran has highly optimized compilers for numerical code and scientific computing development, C++ has much better support for object-oriented programming and a rich set of built-in data structures. In addition, a growing pool of developers have experience with C++, enabling broader participation by the scientific community.

It was decided to perform an essentially direct line-by-line translation of the svFSI solver code into C++ because svFSI had been verified and used successfully for several years. Since then, svMultiPhysics has undergone substantial independent development, and the code has evolved well beyond a line-by-line translation of the original Fortran. This provided an implementation of the core components of the solver; further refactoring into an object-oriented framework is actively underway. Because a line-by-line translation does not by itself guarantee numerical equivalence (Fortran and C++ differ, for example, in array indexing, storage order, argument passing, and default initialization), agreement of svMultiPhysics with published results of its Fortran predecessor is demonstrated in Appendices~\ref{app:fsi-benchmark} and~\ref{app:cardiac-mechanics-benchmark}. During the translation, the C++ results of the integration tests were compared with svFSI reference solutions. Because svMultiPhysics is the direct successor of svFSI, svFSI is not listed separately in Table~\ref{table:comparison}.

The goal of this paper is to provide a comprehensive and accessible reference for svMultiPhysics, including its governing physical models, numerical formulations, software structure, and multiphysics capabilities. We present representative simulations covering computational fluid dynamics (CFD), fluid--structure interaction (FSI), and cardiac electrophysiology to assess the solver's computational performance on physiologically realistic geometries. We further verify the implementation against published benchmarks for incompressible flow past a circular cylinder (Appendix~\ref{app:cfd-benchmark}) and for FSI, cardiac electrophysiology, and cardiac mechanics (Appendices~\ref{app:fsi-benchmark}--\ref{app:cardiac-mechanics-benchmark}). We also demonstrate coupled whole-heart FSI driven by active contraction (Appendix~\ref{app:whole-heart-coupling}) and electromechanics in an idealized left ventricle coupled to a lumped-parameter circulation (Appendix~\ref{app:electromechanics}). Table~\ref{table:comparison} summarizes the capabilities of svMultiPhysics relative to related frameworks. By consolidating the current functionality of svMultiPhysics in a single document, we aim to establish it as a robust, extensible, and community-driven platform for cardiovascular multiphysics modeling.

\begin{table}[!ht]
\begin{adjustwidth}{-0.45in}{-0.45in}
\centering
\caption{{\bf Capabilities of svMultiPhysics and related cardiovascular frameworks.} Based on published capabilities and public documentation as of November 2026; ``--'' denotes not reported.}
\scriptsize
\setlength{\tabcolsep}{2.5pt}
\begin{tabular}{|l|l|l|l|l|l|l|l|}
\hline
 & {\bf svMultiPhysics} & {\bf lifex} & {\bf 4C} & {\bf CHeart} & {\bf IBAMR} & {\bf Alya} & {\bf openCARP} \\ \thickhline
Language & C++ & C++ & C++ & Fortran 2008 & C++ & Fortran & C++ \\ \hline
License & BSD-3 & \begin{tabular}[t]{@{}l@{}}part LGPL-3.0,\\ part in-house\end{tabular} & LGPL & in-house (academic) & \begin{tabular}[t]{@{}l@{}}part BSD-3,\\ part in-house\end{tabular} & proprietary & Academic Public Lic. \\ \hline
Actively developed & yes & yes & yes & yes & yes & yes & yes \\ \hline
Incompressible CFD & yes & yes & yes & yes & yes & yes & -- \\ \hline
Cardiac solid mechanics & yes & yes & yes & yes & yes & yes & -- \\ \hline
FSI & ALE, CMM & ALE & ALE & ALE & IB & ALE & -- \\ \hline
Immersed valve models & RIS/URIS & RIIS & -- & resistive valve planes & IB & -- & -- \\ \hline
Electrophysiology & monodomain & monodomain & monodomain & monodomain & -- & monodomain & mono-/bidomain \\ \hline
CellML import & no & no & -- & via C export & -- & -- & yes \\ \hline
Electromechanics & yes & yes & yes & yes & -- & yes & -- \\ \hline
Implicit 3D--0D coupling & yes & yes & yes & yes & -- & yes & -- \\ \hline
GPU acceleration & linear solvers & no & partial & partial & -- & partial & solvers, ionic models \\ \hline
\end{tabular}
\label{table:comparison}
\end{adjustwidth}
\end{table}

\section*{Solver capabilities and methods}
\subsection*{Governing physics}
This section provides an overview of the governing physics represented in 
svMultiPhysics. The solver is designed as a general multiphysics 
framework and can capture a broad spectrum of phenomena, including fluid 
dynamics, valve dynamics, solid mechanics, scalar transport, and electrophysiology. Covering 
every available model and formulation in detail would go beyond the scope of 
this paper; instead our aim is to give readers a clear sense of the solver's capabilities. The focus here is on presenting the fundamental governing equations of 
the primary physics categories and highlighting the main coupling strategies, 
with particular emphasis on fluid mechanics and fluid-structure interaction.

\subsubsection*{Fluid dynamics}

For cardiovascular applications, the fluid is modeled as incompressible. The governing equations of fluid dynamics are therefore the incompressible Navier-Stokes equations:
\begin{eqnarray}
\nabla \cdot \mathbf{u} &=& 0, \label{eq:continuity} \\[6pt]
\rho \frac{\partial \mathbf{u}}{\partial t} + \rho(\mathbf{u} \cdot \nabla)\mathbf{u} 
&=& \nabla \cdot \boldsymbol{\sigma} + \mathbf{f},
\label{eq:momentum}
\end{eqnarray}
where Equation~\eqref{eq:continuity} enforces mass conservation through the incompressibility constraint, and Equation~\eqref{eq:momentum} represents conservation of momentum. In this system, $\mathbf{u}$ is the velocity vector, $\rho$ the fluid density, $\mathbf{f}$ a body force per unit volume, and $\boldsymbol{\sigma}$ the Cauchy stress tensor defined as
\begin{equation}
    \boldsymbol{\sigma}=-p\mathbf{I}+2\mu(\dot{\gamma})\boldsymbol{\epsilon},
\end{equation}
with $p$ the hydrodynamic pressure,
$\boldsymbol{\epsilon}=\frac{1}{2}(\nabla \mathbf{u}+{\nabla \mathbf{u}}^T)$
the strain-rate tensor, $\dot{\gamma}=\sqrt{2\boldsymbol{\epsilon}:\boldsymbol{\epsilon}}$ the scalar shear rate, and $\mu(\dot{\gamma})$ the dynamic viscosity, which may depend on shear rate to account for non--Newtonian properties of blood. Blood is a complex mixture that consists of plasma, blood cells, and platelets, and blood viscosity is dependent on several factors such as temperature, hematocrit, and, especially, the shear rate. Experimental studies have determined that blood behaves like Newtonian flow at high shear rate, greater than approximately $\qty{100}{\per\second}$~\cite{rahn1967}. In most arteries, such as the aorta and coronary arteries, the shear rate is well above this threshold and blood can be treated as a Newtonian fluid with constant viscosity. On the other hand, when the shear rate is below this threshold, blood exhibits strong shear thinning behavior, i.e. the viscosity decreases with increasing shear rate. Many viscosity models have been proposed to represent this non-Newtonian behavior~\cite{chandran2006}. Currently, svMultiPhysics supports three viscosity models: Newtonian, Carreau-Yasuda, and Casson~\cite{boyd2007}.

The fluid equations can be solved either on a fixed computational domain or, in coupled FSI simulations, on a moving domain. In the arbitrary Lagrangian--Eulerian (ALE) formulation, the fluid mesh deforms with the surrounding structure so that the moving fluid-solid interface is represented explicitly. The residual-based variational multiscale formulation~\cite{hughes2000,bazilevs2007} used by the solver provides a Large Eddy Simulation (LES)-like treatment of unresolved velocity scales, giving users a turbulence-modeling capability within the finite element framework. It is worth noting that svMultiPhysics can also model flows in rigid, stationary porous media. This is accomplished by adding a Darcy resistance term $-\frac{\mu}{K}\mathbf{u}$ to the right-hand side of the momentum equation~\cite{gerosa2024,fuchsberger2022}, where $K$ denotes the permeability of the porous medium.

\subsubsection*{Valve modeling}

A heart valve modeling approach based on the resistive immersed surface (RIS) method \cite{Astorino2012,Fedele2017,This2020,Bucelli2023} is implemented in svMultiPhysics \cite{zhao2026ris}. In this framework, heart valves are represented as immersed surfaces embedded within an unfitted fluid mesh. The valve opening and closing dynamics are prescribed from input data. For an immersed valve surface $\Gamma$ with prescribed motion, its velocity is denoted by $\mathbf{u}_{\Gamma}$. In the RIS formulation, a Darcy-type resistive force is added to the left-hand side of the momentum equation
\begin{equation}
    f_{\Gamma} = R \delta_{\Gamma} (\mathbf{u} - \mathbf{u}_{\Gamma}) \text{ ,}
\end{equation}
where $R$ is the resistance coefficient, and $\delta_{\Gamma}$ is a regularized Dirac delta function that distributes the resistive force from the immersed valve surface to the surrounding fluid domain. This formulation enforces the flow motion in accordance with the prescribed valve motion by penalizing the velocity difference between the fluid and the valve surface.

\subsubsection*{Solid mechanics}
The structural behavior of cardiovascular systems is typically modeled with 
elastic or hyperelastic materials to capture the large deformations of vessel 
walls and cardiac tissue~\cite{fung1993}. The governing equations of solid 
mechanics follow from the balance of linear momentum written in the reference 
configuration:
\begin{eqnarray}
\rho_0 \, \frac{\partial^2 \mathbf{d}}{\partial t^2} &=& \nabla_0 \cdot \mathbf{P} + \mathbf{f}_0,
\label{eq:solid_momentum}
\end{eqnarray}
where $\mathbf{d}$ is the displacement vector, $\rho_0$ the reference density, 
$\mathbf{f}_0$ the body force per unit reference volume, and $\mathbf{P}$ the 
first Piola--Kirchhoff stress tensor. The operator $\nabla_0 \cdot$ denotes the 
divergence with respect to the reference coordinates.

The deformation of the body is described by the deformation gradient
\begin{eqnarray}
\mathbf{F} = \mathbf{I} + \nabla_0 \mathbf{d},
\end{eqnarray}
from which strain measures are derived. A commonly used strain measure is the 
Green--Lagrange strain tensor
\begin{eqnarray}
\mathbf{E} = \tfrac{1}{2}(\mathbf{F}^T\mathbf{F} - \mathbf{I}).
\end{eqnarray}

The material response is expressed through a constitutive relation~\cite{holzapfel2002}. In the
general hyperelastic setting, a strain energy density function 
$W(\mathbf{F})$ is introduced, from which the second Piola--Kirchhoff stress tensor $\mathbf S$ is defined as
\begin{eqnarray}
\mathbf{S} = \frac{\partial W}{\partial \mathbf{E}}.
\end{eqnarray}
The corresponding first Piola--Kirchhoff stress tensor used in the momentum balance is then obtained through
\begin{eqnarray}
\mathbf{P} = \mathbf{F}\mathbf{S}.
\end{eqnarray}

In svMultiPhysics, $\mathbf{S}$ is decomposed into an isochoric (deviatoric), a volumetric, a viscous, and an active contribution,
\begin{eqnarray}
\mathbf{S} &=& \mathbf{S}_{\text{iso}} + \mathbf{S}_{\text{vol}} + \mathbf{S}_{\text{visc}} + \mathbf{S}_{\text{act}},
\end{eqnarray}
The split into isochoric and volumetric terms facilitates modeling nearly incompressible materials, which are common in cardiovascular applications. svMultiPhysics also provides two viscosity models that can be added to the formulation. The current volumetric, isochoric, and viscous constitutive models supported by svMultiPhysics are in Table~\ref{table:constitutive}. The strain energy functions implemented for each model can be found in ~\nameref{S1_Appendix}. 

\begin{table}[!ht]
\centering
\caption{
{\bf Constitutive models supported by svMultiPhysics.}}
\begin{tabular}{l|l}
\hline
\multicolumn{1}{|l|}{\bf Volumetric models} & \multicolumn{1}{|l|}{\bf Isochoric models}\\ \thickhline
Quadratic & Saint Venant-Kirchhoff~\cite{holzapfel2002} \\ \hline
Simo-Taylor91 \cite{simotaylor91} & Modified Saint Venant-Kirchhoff~\cite{holzapfel2002} \\ \hline
Miehe94 \cite{miehe94} & Neo-Hookean~\cite{holzapfel2002} \\ \hline
 & Mooney-Rivlin~\cite{holzapfel2002} \\ \thickcline{1-1}\cline{2-2}
\multicolumn{1}{|l|}{\bf Viscous models} & Holzapfel-Gasser-Ogden~\cite{gasser2006} \\ \thickcline{1-1}\cline{2-2}
Newtonian~\cite{ortiz1999} & Guccione~\cite{guccione1991} \\ \hline
Potential~\cite{chapelle2012} & Holzapfel-Ogden~\cite{holzapfel2009} \\ \hline
 & Holzapfel-Ogden Modified Anisotropy~\cite{nolan2014,shi2024} \\ \hline
 & Custom invariant-based material model \\
 & (constitutive artificial neural network framework)~\cite{Peirlinck2025} \\ \hline
\end{tabular}
\label{table:constitutive}
\end{table}

For active materials, the active stress contribution is prescribed using the orthonormal basis defined by the fiber, sheet, and sheet-normal directions, denoted by $\mathbf f$, $\mathbf s$, and $\mathbf n$, respectively:
\begin{equation}
\mathbf S_{\text{act}} =
T_a \left(
\eta_f \mathbf f \otimes \mathbf f +
\eta_s \mathbf s \otimes \mathbf s +
\eta_n \mathbf n \otimes \mathbf n
\right),
\label{eq:active_stress}
\end{equation}
where $T_a$ denotes the active tension magnitude and $\eta_f$, $\eta_s$, and $\eta_n$ are non-negative weights distributing it along the fiber, sheet, and sheet-normal directions. svMultiPhysics provides two ways to define $T_a$. In purely mechanical simulations, $T_a$ can be prescribed as a spatially uniform function of time, either constant or a user-defined waveform (e.g., the double-Hill function used in Appendix~\ref{app:whole-heart-coupling}). In electromechanical simulations, $T_a$ is computed at every node from the electrophysiology solution through an excitation--contraction model driven by the intracellular calcium concentration $c=[\mathrm{Ca}^{2+}]_i$ supplied by the ionic model (for phenomenological ionic models, by a calcium proxy variable). Three such models are available. The first is of Nash--Panfilov type~\cite{nashpanfilov2004,goktepe2010},
\begin{equation}
\frac{dT_a}{dt}=\varepsilon(c)\left[\eta_T\,(c-c_{\mathrm{rest}})-T_a\right],\qquad
\varepsilon(c)=\varepsilon_0+(\varepsilon_i-\varepsilon_0)\exp\!\left(-e^{-\xi_T(c-c_{\mathrm{crit}})}\right),
\label{eq:nashpanfilov}
\end{equation}
in which the transmembrane potential of the original formulation is replaced by $c$. The second is the biophysically detailed RDQ20-MF model of sarcomere force generation~\cite{regazzoni2020}, in which $T_a$ depends on $c$ and on the sarcomere length computed from the local fiber stretch. The third is the phenomenological-driven Land--Niederer model of human cardiomyocyte contraction~\cite{land2017}, in which $T_a$ depends on the calcium--troponin binding, the cross-bridge states, and the fiber stretch. In the current implementation, the fiber-stretch-rate (force--velocity) feedback of the published RDQ20-MF and Land--Niederer models is neglected, so the implemented models differ slightly from their published versions. In all three cases, the active-tension equations are integrated pointwise at each time step and the resulting $T_a$ enters Eq~(\ref{eq:active_stress}). Appendix~\ref{app:electromechanics} demonstrates the RDQ20-MF model driven by the ten Tusscher--Panfilov calcium transient in an idealized left ventricle. svMultiPhysics provides Python scripts to generate the vectors $\mathbf f$, $\mathbf s$, and $\mathbf n$ using Laplace-Dirichlet rule-based fiber fields following the methods presented in Bayer et al. (2012) \cite{bayer2012} or Doste et al. (2019) \cite{doste2019}.

The svMultiPhysics solver builds upon this general formulation to provide
several solid mechanics capabilities:
\begin{itemize}
  \item Linear elastodynamics: solves the small-deformation linear
  elasticity problem.
  \item Linear elastodynamics for mesh motion: solves for
  elastic deformation of the mesh in fluid--structure interaction simulations,
  ensuring a smooth propagation of boundary motion.
  \item Nonlinear elastodynamics: solves the nonlinear structural
  mechanics problem, suitable for
  hyperelastic models of cardiovascular tissue. Two formulations are available: 1) a standard displacement-based formulation, 2) a mixed pressure-velocity formulation, also referred to as the \texttt{ustruct} formulation, based on the Gibbs free energy. In this formulation, 
  \begin{equation}
      \mathbf S_{\text{vol}} = -p J (\mathbf F^T \mathbf F)^{-1},
  \end{equation}
  where $p$ is the hydrostatic pressure and $J=\det \mathbf F$.
  The mixed formulation is advantageous for nearly incompressible soft tissues and provides the solid mechanics component of the unified VMS framework, allowing fluid, solid, and FSI problems to be treated consistently within the same numerical formulation~\cite{liu2018}.
  \item Nonlinear thin shell mechanics: models nonlinear shell
  mechanics using Kirchhoff--Love shell theory, allowing efficient simulation
  of thin-walled structures such as heart valves or vascular grafts. The
  shell is represented by its midsurface, with displacement degrees of freedom
  defined on that surface and transverse shear deformation neglected, which is
  appropriate for thin shells. In weak form, the Kirchhoff--Love shell problem
  can be written as
  \begin{equation}
      \int_{\Gamma_0} \left( \mathbf{n}^{\alpha\beta}\,\delta \boldsymbol{\varepsilon}_{\alpha\beta}
      + \mathbf{m}^{\alpha\beta}\,\delta \boldsymbol{\kappa}_{\alpha\beta} \right)\,dA
      = \int_{\Gamma_0} \mathbf f \cdot \delta\mathbf d\,dA
      + \int_{\partial\Gamma_0} \mathbf t \cdot \delta\mathbf{d} \,ds,
  \end{equation}
  where $\Gamma_0$ is the reference midsurface, $\mathbf{d}$ is the midsurface
  displacement, $\mathbf{n}^{\alpha\beta}$ and $\mathbf{m}^{\alpha\beta}$ are the membrane force
  and bending moment resultants, $\boldsymbol{\varepsilon}_{\alpha\beta}$ and
  $\boldsymbol{\kappa}_{\alpha\beta}$ are the midsurface membrane strain and curvature
  change tensors, and $\mathbf f$ and $\mathbf t$ denote prescribed surface and
  boundary tractions, respectively.
\end{itemize}

\subsubsection*{Scalar advection--diffusion}
Advection--diffusion equations describe the transport of a scalar quantity within a medium due to the combined effects of advection and diffusion. Common 
applications include modeling temperature distributions in solids or fluids, or 
tracking the concentration of a passive scalar (e.g., a dye or chemical species) 
carried by a flow field. The general unsteady advection--diffusion equation in 
conservative form is written as
\begin{eqnarray}
\frac{\partial g}{\partial t} + \nabla \cdot (\mathbf{u} g) &=& 
\nabla \cdot (k \nabla g) + f,
\label{eq:advectiondiffusion}
\end{eqnarray}
where $g$ denotes the scalar of interest, $\mathbf{u}$ is a prescribed velocity 
field, $f$ is a source or sink term, and $k$ is the diffusivity of the medium. 
The diffusivity $k$ can be a scalar, in which case diffusion is isotropic and 
uniform in all directions, or a tensor, in which case diffusion is anisotropic 
and directionally dependent. For the common case of constant isotropic 
diffusivity, the diffusion operator simplifies to $k \nabla^2 g$. 

In cardiovascular flow simulations, this scalar transport framework can also be used to compute hemodynamic residence time in an Eulerian manner. Rather than releasing and tracking discrete particles, residence time can be estimated by solving an auxiliary advection--diffusion problem for a passive scalar over a region of interest, providing a continuous measure of flow stasis that is useful for assessing recirculation and potential thrombotic risk~\cite{Esmaily2013ResidenceTime}.

\subsubsection*{Cardiac electrophysiology}
The electrical activity of the heart originates in the sinoatrial node, a small cluster of pacemaker cells located in the right atrium. These cells periodically emit electrical impulses that propagate rapidly through the heart’s conduction system, ensuring coordinated activation. The electrical signal travels from the sinoatrial node to the atrioventricular node and then through the Purkinje fiber network, which connects to the ventricular myocardium via Purkinje--myocardium junctions, triggering a nearly simultaneous contraction of the ventricles.
On the cellular level, this propagation is mediated by depolarization: voltage-gated ion channels in the cell membrane open, allowing positively charged ions to move across the membrane. The resulting ionic currents generate action potentials that spread between neighboring cells, forming the basis of coordinated electrical activation in the heart.
A commonly used model for the propagation of cardiac electrical activity is the monodomain equation, which takes the form of a reaction–diffusion system:
\begin{eqnarray}
    \frac{\partial V}{\partial t} + \frac{I_{ion}(V, \mathbf{w}) - I_{app}(t)}{C_m}  = \nabla \cdot ( \mathbf{D}\nabla V)
\label{eq:propagation}
\end{eqnarray}
where $V$ is the trans-membrane potential and $C_m$ the membrane capacitance per unit area. The terms $I_{ion}$ and $I_{app}$ represent the ionic current density (per unit area) and externally applied current, respectively. Here $\mathbf{D}$ is a diffusivity tensor, with units of length squared per unit time. Writing the dimensional monodomain equation as $\chi C_m\,\partial V/\partial t+\chi(I_{ion}-I_{app})=\nabla\cdot(\boldsymbol{\sigma}_m\nabla V)$ and dividing by $\chi C_m$ gives Eq~(\ref{eq:propagation}) with $\mathbf{D}=\boldsymbol{\sigma}_m/(\chi C_m)$, for spatially constant $\chi C_m$. Thus the membrane surface-to-volume ratio $\chi$ is absorbed into $\mathbf{D}$; a volumetric stimulus is divided by $\chi$ to express it per unit membrane area. The conductivity-to-diffusivity conversion is illustrated in Appendix~\ref{app:ep-benchmark}. The diffusion tensor $\mathbf{D}$ controls the spread of the electrical signal through the tissue; in isotropic tissue, it reduces to a scalar diffusivity, while in anisotropic myocardium it reflects the preferential propagation along fiber directions, commonly assuming the conductivity is transversely isotropic. As with the cardiac mechanics, the fiber direction can be generated using custom Python scripts available in svMultiPhysics. The diffusivity tensor is defined using isotropic and orthotropic diffusivity coefficients, $D_{\text{iso}}$ and $D_{\text{f}}, D_{\text{s}}, D_{\text{n}}$, respectively, such that
\begin{equation}
\mathbf{D} = D_{\text{iso}} \mathbf{I} + D_{\text{f}} \, \mathbf{f} \otimes \mathbf{f} + D_{\text{s}} \, \mathbf{s} \otimes \mathbf{s} + D_{\text{n}} \, \mathbf{n} \otimes \mathbf{n},
\end{equation}
where $\mathbf{f}, \mathbf{s}, \mathbf{n}$ denote the unit vectors describing the fiber, sheet, and sheet normal orientations defined over the computational domain.
The ionic current term $I_{ion}$ encapsulates the dynamics of depolarization and repolarization at the cellular level. Depending on how these processes are represented, ionic models can be broadly divided into two categories: biophysics-based ionic models, which describe ion-channel kinetics in detail, and phenomenological models, which capture the essential features of action potential generation and propagation with reduced complexity. The available cardiac ionic models in svMultiPhysics are listed in Table~\ref{table:electrophysiology}.
\begin{table}[!ht]
\centering
\caption{
{\bf Cardiac ionic models supported by svMultiPhysics.}}
\begin{tabular}{l|l}
\hline
\multicolumn{1}{|l|}{\bf Biophysics-based models} & \multicolumn{1}{|l|}{\bf Phenomenological models}\\ \thickhline
ten Tusscher--Panfilov~\cite{tentusscher2004,tentusscher2006} & Aliev--Panfilov~\cite{aliev1996,goktepe2009} \\ \hline
 & FitzHugh--Nagumo~\cite{fitzhugh1961,nagumo1962,goktepe2009} \\ \hline
 & Bueno-Orovio-Cherry-Fenton~\cite{bueno2008} \\ \hline
\end{tabular}
\label{table:electrophysiology}
\end{table}

The electrophysiology implementation supports 1D, 2D, and 3D geometries and enables coupling between computational domains of different topologies at user-specified nodes. The coupling is established by identifying nearest-neighbor connections between meshes, allowing electrical signals to propagate across domains. This capability is particularly useful for modeling the propagation of the transmembrane potential from a 1D representation of the Purkinje network to the 3D ventricular myocardium through Purkinje--myocardium junctions.
Electrocardiograms can be calculated for an arbitrary number of electrodes in svMultiPhysics by providing a list of the electrode coordinates $\mathbf x_e$ using \cite{salvador2024},
\begin{equation}
    \phi_e(\mathbf x_e) = -\int_{\Omega} \nabla V \cdot \nabla \frac{1}{\| \mathbf x - \mathbf x_e \|} \;dV,
    \label{eq:ecg}
\end{equation}
where $\phi_e$ is the electrocardiogram signal for electrode $e$. Eq~(~\ref{eq:ecg}) corresponds to the classical pseudo-ECG of Gima and Rudy~\cite{gima2002}, $\phi_e=\frac{a^2\sigma_i}{4\sigma_e}\int_\Omega(-\nabla V)\cdot\nabla\frac{1}{\|\mathbf x-\mathbf x_e\|}\,dV$, which assumes an unbounded, homogeneous, and isotropic volume conductor, with the constant prefactor $a^2\sigma_i/(4\sigma_e)$ set to unity. Consequently, no extracellular conductivity needs to be prescribed, and the computed signals are in arbitrary units: ECG morphology and timing are preserved, whereas absolute amplitudes require scaling by this prefactor. Torso inhomogeneities and intracellular anisotropy are not represented in the ECG computation.

\subsection*{Numerical methods}
\subsubsection*{Discretization methods}
The partial differential equations (PDEs) governing the various physical phenomena modeled in svMultiPhysics are discretized in space using the finite element method (FEM)~\cite{hughes1987}. svMultiPhysics supports continuous finite elements, including linear tetrahedra, prisms, and hexahedra. Support is not restricted to linear elements: the FSI benchmark uses biquadratic elements (Appendix~\ref{app:fsi-benchmark}), and the cardiac mechanics benchmark uses quadratic tetrahedra (Appendix~\ref{app:cardiac-mechanics-benchmark}). Originally developed for solid mechanics applications, the FEM relies on the Galerkin method for discretizing the weak form of the governing equations. When extended to computational fluid dynamics (CFD), however, the standard Galerkin formulation was found to be unstable due to the presence of nonlinear convective terms in the fluid equations. This limitation led to the development of stabilized FEM formulations, such as the streamline upwind/Petrov–Galerkin (SUPG) and related methods~\cite{tabata1977,hughes1984,brooks1982}. Building on these ideas, svMultiPhysics adopts the variational multiscale (VMS) framework, which can be interpreted as an LES-like formulation~\cite{hughes2000}. The governing equations for fluid dynamics are therefore solved using a residual-based VMS finite element formulation~\cite{bazilevs2007}. Moreover, svMultiPhysics offers the possibility to use a unified VMS formulation that consistently treats fluid, solid, and fluid–structure interaction (FSI) problems within the same numerical framework~\cite{liu2018}. A broader review of multiscale cardiovascular flow modeling, including stabilized discretizations, coupled boundary conditions, and solver strategies relevant to patient-specific simulations, is provided in~\cite{Marsden2015Multiscale}.

The unsteady governing equations are discretized in time with the generalized-$\alpha$ method, an implicit second order accurate scheme developed for stabilized finite element formulations ~\cite{jansen2000}. Based on previous work~\cite{whiting2001,bazilevs2007}, a one-parameter family of unconditionally stable time integrators can be obtained by varying the spectral radius, $\rho_{\infty}$, of the amplification matrix at an infinite time step. The $\rho_{\infty}$ parameter, controls high-frequency damping with respect to the temporal resolution of the problem and it can take values between $0$ and $1$, where $0$ represents maximum damping and $1$ means all frequencies are retained~\cite{chung1993}.

The non-linear governing equations are solved using the Newton-Raphson method. As a result, a final linear system of equations is obtained, which is solved at each Newton iteration of each timestep. The svMultiPhysics solver interfaces with different linear algebra packages to solve the linear systems as described later.

\subsubsection*{Boundary conditions}
The governing equations for fluid dynamics, scalar advection-diffusion, and solid mechanics are formulated as initial-boundary-value problems, which require appropriate initial conditions and boundary conditions on the computational domain in order to be well-posed. svMultiPhysics supports three classical types of boundary conditions: Dirichlet, Neumann, and Robin, as well as coupled boundary conditions that link the three-dimensional computational domain to reduced-order ``lumped parameter'' representations of the heart and circulatory system~\cite{Esmaily2013CoupledBC,Brown2024ModularCoupling,moghadam2013}.
Dirichlet boundary conditions prescribe the value of the solution variables directly, for example, velocity on inflow boundaries in fluid problems or displacement in solid mechanics.
Neumann boundary conditions prescribe the normal traction, such as pressure on outflow boundaries in fluid problems.
Robin boundary conditions specify a mixed condition, and they apply a traction analogous to a spring-damper system.
Boundary condition data may be constant, spatially varying, or time-dependent, allowing users to define realistic physiological or mechanical constraints on any portion of the boundary.
Coupled boundary conditions are achieved via specialized lumped-parameter boundary conditions that model the downstream vasculature and circulatory physiology. These models represent physiologic impedance using analogies to electrical circuits~\cite{kung2013,seo2020,arthurs2020}. A simple choice is a pure resistance boundary condition, while the most widely used model in cardiovascular flows is the three-element Windkessel, or RCR model, which combines proximal resistance, compliance, and distal resistance~\cite{vignon2010,vignon2006}.
In addition, svMultiPhysics provides an interface to the svZeroDSolver as a plugin (see \nameref{sv0d}), enabling users to prescribe general lumped-parameter networks (LPNs) as boundary conditions. In these 3D--0D formulations, the finite element model supplies interface quantities such as flow rate or pressure to a reduced-order circuit model, while the 0D model returns the corresponding boundary response imposed on the 3D domain. This coupling can be applied to blood flow problems, where LPNs represent downstream vascular impedance, and to solid mechanics problems, where reduced-order models can represent circulatory or tissue-level interactions. Implicit coupling and appropriate preconditioning are important for stability and efficiency when the 3D and 0D systems are strongly coupled~\cite{Esmaily2013CoupledBC,Brown2024ModularCoupling,moghadam2013}.  This avoids the need for explicitly coupled systems often used in general solvers, which impose severe time step restrictions to maintain stability.  

\subsubsection*{Linear algebra}
The svMultiPhysics solver interfaces with three linear algebra backends: the in-house library FSILS, and the external libraries Trilinos and PETSc, thus providing access to a wide range of linear solvers and preconditioners.
A summary of all available linear solvers (LS) and preconditioners (PC) is presented in Table~\ref{table:linearalgebra}.
\begin{table}[!ht]
\centering
\caption{
{\bf Linear algebra available in svMultiPhysics.}}
\begin{tabular}{ll|ll|ll}
\hline
\multicolumn{2}{|l|}{\bf FSILS} & \multicolumn{2}{l|}{\bf Trilinos} & \multicolumn{2}{|l|}{\bf PETSc} \\ \thickhline
\bf LS         & \bf PC            & \bf LS         & \bf PC               & \bf LS         & \bf PC           \\ \hline
BIPN      & Jacobi      & GMRES      & Jacobi         & GMRES      & Jacobi     \\
GMRES       & RCS           & CG         & AMG               & CG         & RCS          \\
CG         &               & BICGS      & ILU              & BICGS      &              \\
BICGS      &               &            & ILUT             &            &              \\
           &               &            & Block Jacobi     &            &              \\
           &               &            & RILU             &            &              \\ \hline
\end{tabular}
\label{table:linearalgebra}
\end{table}
The Krylov-based iterative solvers implemented in svMultiPhysics include the generalized minimal residual (GMRES) method, the conjugate gradient (CG) method, the biconjugate gradient stabilized (BICGS) method, and the bi-partitioned iterative algorithm (BIPN)~\cite{moghadam2015}. 

The BIPN method is particularly efficient for solving the linear systems arising from the incompressible Navier–Stokes equations, as it decouples the pressure and velocity blocks of the system matrix to accelerate convergence. In svMultiPhysics, the BIPN solver is paired with a resistance-based preconditioner, specifically designed to improve the conditioning of matrices encountered in cardiovascular simulations~\cite{moghadam2013}, where resistance or RCR boundary conditions often lead to significant matrix ill-conditioning~\cite{vignon2010}. The preconditioner incorporates the dominant contribution of the outlet resistance terms into the approximate linear solve, rather than treating these strongly coupled boundary effects only through the original system matrix. By accounting for the impedance imposed by downstream vascular models, the method reduces the ill-conditioning associated with large resistance values and improves Krylov convergence for hemodynamic flow problems. Although highly efficient for fluid dynamics applications, this implementation is specialized for CFD problems. The resistance-based preconditioner is implemented within the Jacobi preconditioning framework since it naturally reduces to a standard Jacobi preconditioner in the absence of resistances. Presently in svMultiPhysics, the BIPN solver inherently operates with the resistance-based preconditioner; using the solver without it is not available. For broader physics applications, svMultiPhysics leverages the extensive preconditioning capabilities available through Trilinos and PETSc, including ILU, Jacobi, block-Jacobi, row-column scaling (RCS) and algebraic multigrid (AMG) schemes, thereby enhancing solver robustness and performance across diverse problem types.

When Trilinos is configured with device-enabled Tpetra~\cite{baker2012tpetra} and Kokkos~\cite{trott2021kokkos} backends, the Krylov solvers and associated preconditioners execute on GPUs. In this configuration, sparse matrix-vector products, vector updates, and preconditioner applications are offloaded to the device (GPU), while the higher-level nonlinear iteration and finite element assembly remain on the host (CPU). This hybrid CPU-GPU strategy provides significant acceleration for large-scale problems where the linear solve dominates computational cost.

\subsubsection*{Coupling strategies}
svMultiPhysics is designed to support strongly coupled multiphysics formulations. In particular, the primary FSI implementation is monolithic: the fluid and structural unknowns are assembled and solved together within a single nonlinear system, rather than advanced through a staggered scheme. Sequential solves are used only for selected auxiliary components or workflows, such as mesh motion in the ALE formulation, where they complement the monolithic coupled solve while preserving flexibility across different physics modules.
For fluid–structure interaction, svMultiPhysics provides two alternative formulations: the Arbitrary Lagrangian–Eulerian (ALE) method and the Coupled Momentum Method (CMM).
The ALE method, originally formulated within the finite element framework by Hughes et al.~\cite{hughes1981ale} and later extended to the residual-based variational multiscale (VMS) formulation by Bazilevs et al.~\cite{bazilevs2012}, is implemented in svMultiPhysics for high-fidelity FSI simulations. In this formulation, the computational fluid mesh moves to conform to the deformation of the solid domain, ensuring an accurate representation of the moving interface. The mesh motion is governed by a linear elastodynamics equation, solved sequentially after the monolithic FSI system.
svMultiPhysics adopts a strongly coupled monolithic ALE formulation, in which the fluid and solid equations are solved simultaneously within a single linear system at each Newton iteration. This approach provides excellent numerical stability and accuracy, particularly for problems involving large structural deformations, strong coupling effects, or incompressible flows with deformable walls, such as the case in cardiovascular applications.
As a computationally cheaper alternative, the Coupled Momentum Method (CMM)~\cite{figueroa2006} is also available. This method assumes a thin-wall structural model, in which the vessel wall is treated as a thin, linearly elastic membrane represented on the two-dimensional lumen surface, enhanced with transverse shear~\cite{figueroa2006}, that interacts with the surrounding fluid through a momentum exchange at the interface. The CMM enables the simulation of wall compliance and wave propagation effects without explicitly resolving the full three-dimensional structure. While restricted to problems with small ($<$10\%) deformations, it offers a significant reduction in computational cost, making it particularly suitable for large-scale cardiovascular simulations or parameter studies where efficiency is critical.

\section*{Software development platform}
svMultiPhysics software is hosted on GitHub (see \nameref{svgithub}) as an open-source software development project. The svMultiPhysics C++ code and associated files (e.g. data for testing) are stored in a repository that can be downloaded using the Git version control system. The software is released under a standard BSD license, making it freely available for both academic and commercial use.

The main components of the repository are the following.
\begin{itemize}
\item Source code
\item Doxygen documentation
\item Integration tests
\item Unit tests
\item Automated workflows for continuous integration/continuous deployment (CI/CD)
\end{itemize}

The CI/CD pipeline, implemented using the GitHub Actions platform, plays a central role in maintaining code quality and robustness as new features are introduced. 
Integration tests verify that benchmark problems covering all supported physics and solver functionalities run successfully and exhibit consistent convergence behavior.
In contrast, unit tests isolate and validate the correctness of individual functions or modules, ensuring that new implementations behave predictably.
Together, this automated testing framework enables new contributions to be merged into the main code base with significantly greater confidence compared to manual integration.
Finally, as an open-source project, svMultiPhysics welcomes contributions from the community. Developers and users are encouraged to follow the contribution guidelines provided in the repository and to report issues or feature requests through the GitHub Issues page.

\subsection*{Building svMultiPhysics}
svMultiPhysics is an MPI--based solver and therefore requires a working MPI toolchain and compatible MPI compilers to build. In addition, the solver uses the Visualization Toolkit (VTK) ~\cite{vtkBook} file formats to import meshes and store simulation results. As a result, VTK is an essential dependency even for a minimal installation.
To enable the full linear algebra capabilities of svMultiPhysics, including the complete set of iterative solvers and preconditioners, the libraries Trilinos and PETSc must also be installed and linked during compilation. Detailed build instructions, including the required CMake configuration, compiler flags, and dependency setup, are provided on the project’s GitHub page (\nameref{svgithub}).
svMultiPhysics can optionally be built with GPU acceleration when linked with a device-enabled Trilinos installation. In this case, Trilinos must be compiled with Kokkos support for the target architecture (e.g. CUDA). When configured properly, the linear solver infrastructure automatically utilizes GPU execution spaces for sparse linear algebra operations. 
Building a research--grade multiphysics solver with multiple third--party dependencies can be challenging, particularly on HPC systems where environment configuration and library compatibility differ across platforms. To address this, a Docker--based build environment has been created and integrated directly into the project’s CI workflow.
This containerization approach serves two distinct purposes:
\begin{itemize}
    \item For users, the Docker image provides a fully configured environment that can be pulled to workstations or HPC systems (via Docker, Singularity, or Apptainer), ensuring a straightforward and reproducible installation with all required dependencies.
    \item For continuous integration, the Docker container is essential for running the automated integration tests. Since these tests require functional builds of VTK, Trilinos, and PETSc, recreating the entire environment from scratch on every GitHub pull request would be prohibitively expensive and, in some cases, infeasible on GitHub Actions. The pre--built container ensures that the CI pipeline has immediate access to a stable and consistent environment, enabling reliable compilation and execution of the full solver test suite, including those cases that depend on Trilinos and PETSc.
\end{itemize}
The corresponding Docker images are hosted on the SimVascular DockerHub repository (see \nameref{svdockerhub}). Each image provides a fully configured environment with all required dependencies, and can also be used to build custom development versions of svMultiPhysics.

\subsection*{Documentation and input structure}
svMultiPhysics documentation is available on the SimVascular website (see \nameref{svdoc}). The guide includes both a practical description of the solver interface and an overview of the theoretical background underlying the physical models and numerical methods implemented in the code. It also provides detailed explanations of the input options, simulation workflow, and available solver functionalities.
svMultiPhysics reads simulation parameters from an Extensible Markup Language (XML) file. The XML format is used to logically structure all solver parameters and input data, such as mesh files, boundary conditions, solver settings, and medium properties, ensuring a clear description of each simulation setup.
A Developer Guide is also provided, focusing on internal code organization, data structures, and implementation details. Its goal is to assist contributors in understanding the architecture of svMultiPhysics and to facilitate future development of the solver. Most integration test cases in the repository (\texttt{tests/cases/<physics>/<case>}) are accompanied by a README describing the problem, the relevant XML input options, and the expected results, and therefore also serve as minimal, physics-specific examples for new users. The status of public input files for the examples and appendix benchmarks is described in the Data availability section.

\subsubsection*{Pre- and post-processing workflow}
In addition to the core solver, svMultiPhysics is supported by a collection of pre- and post-processing workflows that help users prepare models, define simulation inputs, and analyze results. For cardiovascular applications, geometric models and computational meshes are commonly generated within the SimVascular pipeline~\cite{simvascular}, while repository utilities and user scripts can be used to prepare mesh and boundary files, assign fiber directions for electrophysiology and anisotropic material models ($\mathbf f, \mathbf s, \mathbf n$ vectors), convert data formats, and assemble case-specific input files. These tools are intended to complement the XML-based solver interface by reducing the amount of manual data preparation required for complex multiphysics simulations.

Simulation results are written primarily using VTK-based formats, including mesh and boundary data in \texttt{.vtu} files, which enables direct visualization in ParaView and further analysis with Python-based tools such as PyVista. Depending on the selected physics, output fields may include fluid velocity and pressure, solid displacement and stress measures, wall-related quantities, scalar transport variables, electrophysiology variables, and derived quantities used for model assessment. This common output structure provides a flexible basis for comparing simulations, extracting quantitative metrics, and developing custom analysis scripts for application-specific studies.

\section*{Results and discussion}

To demonstrate the use of svMultiPhysics, we present representative cardiovascular simulations that exercise key solver capabilities. The examples are not intended to exhaustively cover all supported physics, but rather to highlight how the same framework can be used for fluid dynamics and fully coupled FSI problems on clinically relevant geometries. We also use these cases to evaluate the effect of different linear solver and preconditioner choices on parallel execution.

Unless stated otherwise, the results in this paper were computed with the svMultiPhysics release v1.0.0.2 (\url{https://github.com/SimVascular/svMultiPhysics/releases/tag/v1.0.0.2}, commit \texttt{fba504e}). This applies to the AAA CFD and FSI CPU performance study, the biventricular electrophysiology study, and the benchmarks in Appendices~\ref{app:cfd-benchmark}--\ref{app:cardiac-mechanics-benchmark}. There are three exceptions: the preliminary CPU--GPU comparison was performed with an earlier development version; the stage-level timing breakdown of the AAA runs was obtained with a local development build that adds per-stage timers, a capability that will be added to svMultiPhysics as a profiling utility (\href{https://github.com/SimVascular/svMultiPhysics/issues/730}{issue \#730}); and the whole-heart example in Appendix~\ref{app:whole-heart-coupling} was computed with an earlier version of svMultiPhysics.

\subsection*{Abdominal aortic aneurysm (AAA) model description}
A 3D model was constructed from a static Computed Tomography (CT) image of a 73-year-old male with an abdominal aortic aneurysm (AAA) \cite{CT_Image_patient_Ortiz2025-bh}. Image segmentation and model generation were performed in SimVascular \cite{simvascular}. To balance physiological fidelity with computational cost, smaller distal branches were removed, resulting in a one-inlet, two-outlet geometry representative of typical clinical anatomies. The lumen volume was meshed using TetGen \cite{TetGen_Si2008-lu} within SimVascular with a six-element boundary layer and approximately $1.5$ million tetrahedral elements and 270,000 nodes. The vessel wall was generated by extruding the fluid–structure interface using PyVista \cite{pyvista_sullivan2019pyvista}, assuming a uniform thickness of $2$ \si[per-mode=symbol]{\milli\meter} based on literature values \cite{FSI_aortic_diss_prestress_Baumler2020-ap}. The wall mesh was then remeshed while preserving the interface surface and contains roughly $340,000$ tetrahedral elements and 80,000 nodes.

\begin{figure}
    \centering
    \begin{minipage}{0.32\linewidth}
        \centering
        \includegraphics[width=\linewidth]{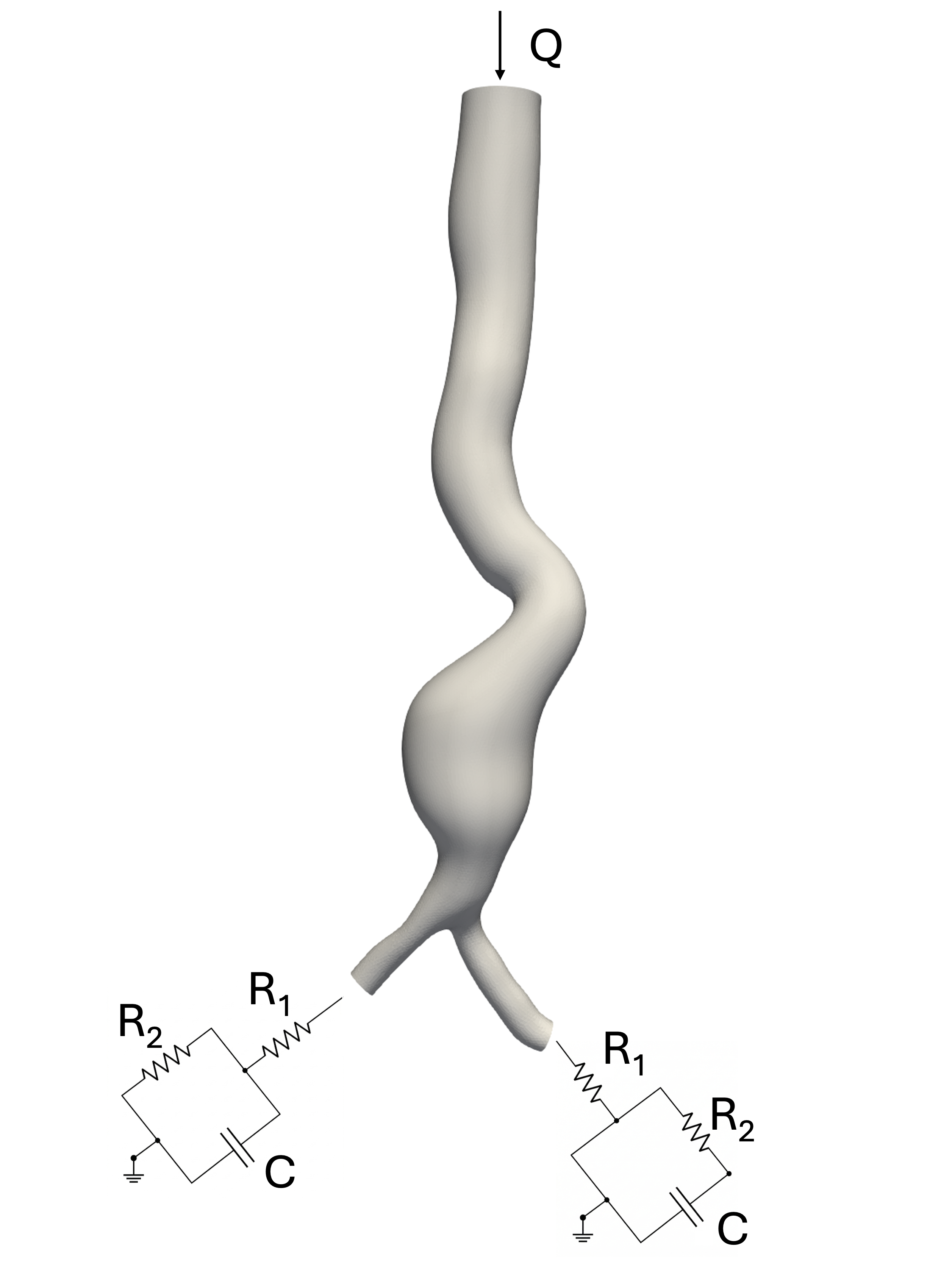}
        \textbf{A}
    \end{minipage}
    \hfill
    \begin{minipage}{0.32\linewidth}
        \centering
        \includegraphics[width=\linewidth]{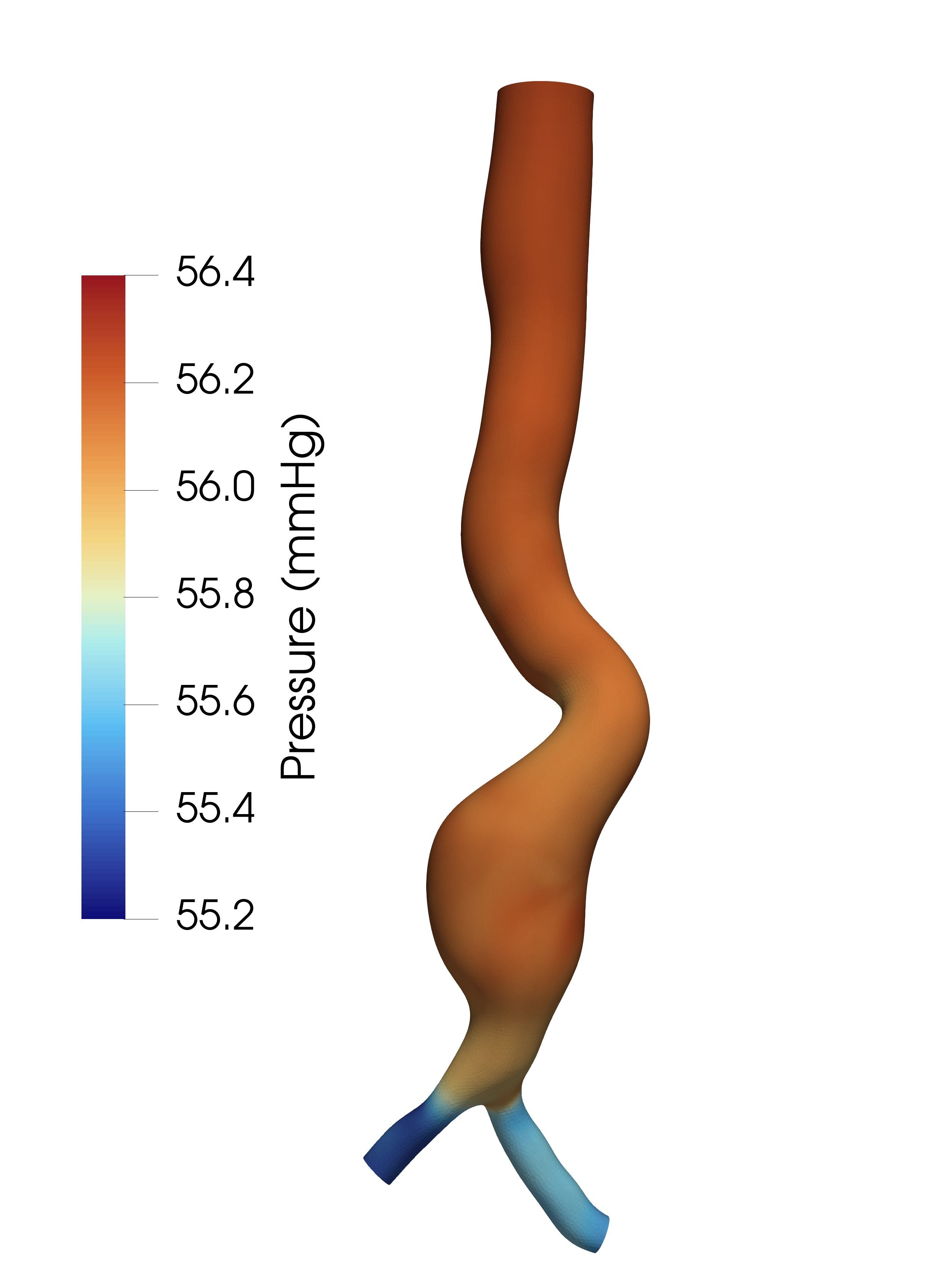}
        \textbf{B}
    \end{minipage}
    \hfill
    \begin{minipage}{0.32\linewidth}
        \centering
        \includegraphics[width=\linewidth]{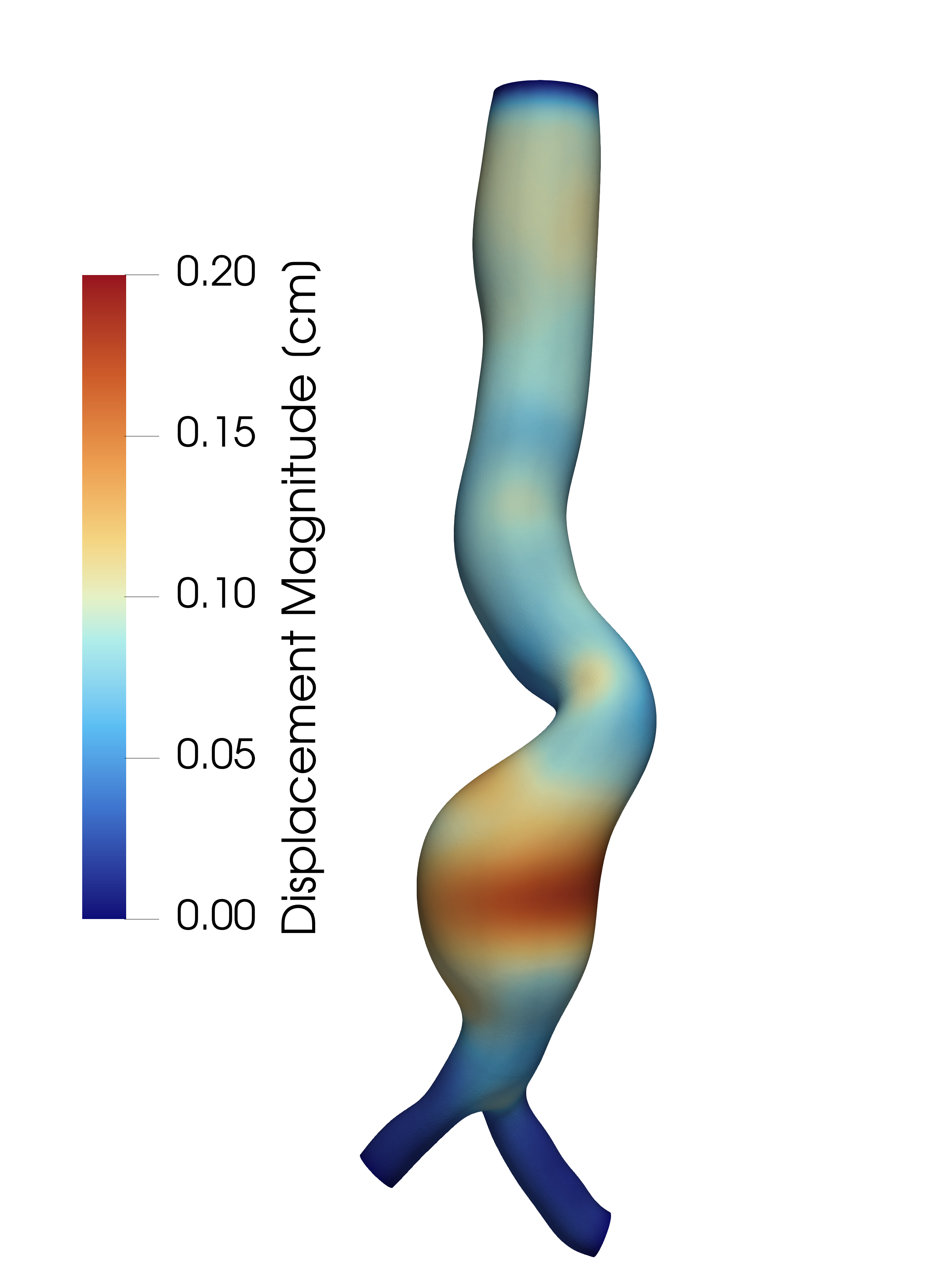}
        \textbf{C}
    \end{minipage}
    \caption{{\bf AAA model and representative pressure fields.}
    (A) Computational domain for the abdominal aortic aneurysm (AAA) model. (B) Pressure field from the CFD simulation at a representative time point after two cardiac cycles; the outlet pressures have not yet reached a periodic state. (C) Displacement magnitude from the ALE-FSI simulation at the same representative time point.}
    \label{fig:aaaDomain}
\end{figure}

At the aortic inlet, a normalized flow waveform\cite{Inflow_Lan2022-zo} was prescribed. Each iliac outlet was coupled to a three-element Windkessel (RCR) model representing the downstream vasculature \cite{RCR_Pfaller2021-oa}. The RCR model includes proximal and distal resistances, representing viscous losses, and a compliance element representing distal vessel elasticity. The RCR parameters are tuned to patient metrics of blood pressure cuff measurements and flow rates from phase contrast MRI. These values define the target pressures of a periodic state; they are not the pressures reached in the short simulations reported here (see below). The mean arterial pressure is calculated as a weighted average of the systolic ($122$ \si[per-mode=symbol]{\mmHg}) and diastolic pressures ($74$ \si[per-mode=symbol]{\mmHg}) yielding $90$ \si[per-mode=symbol]{\mmHg} ($119,990$ \si[per-mode=symbol]{\dyne\per\centi\meter\tothe{2}}). The average inlet flow rate for the patient is $62.79$ \si[per-mode=symbol]{\centi\meter\tothe{3}\per\second} which corresponds to an inlet Reynolds number of $\approx 960$ . The total resistance for each outlet is calculated as

\begin{equation}
    R_{total} = \frac{2P_{mean}}{Q_{mean}}
\end{equation}

\noindent giving $3821.9$ \si[per-mode=symbol]{\dyne\second\per\centi\meter\tothe{5}}. Using a proximal to distal resistance ratio for the abdominal aorta of $0.1$ \cite{RCR_Res_ratio_Maher2021-ir}, the proximal and distal resistances were $R_p = 347.44$ \si[per-mode=symbol]{\dyne\second\per\centi\meter\tothe{5}} and  $R_d = 3474.45$ \si[per-mode=symbol]{\dyne\second\per\centi\meter\tothe{5}}, respectively. The capacitance is estimated as $ C = 4.98 \times 10^{-4}$ \si[per-mode=symbol]{\centi\meter\tothe{5}\per\dyne} following the methodology in \cite{RCR_Kaiser2021-hv} using the pressure waveform at the start and end of the diastolic phase. A Robin boundary condition was applied to the exterior vessel wall to model elastic support from the surrounding tissue, with stiffness coefficient $k_s = 1 \times 10^7~\si{\dyne\per\centi\meter\cubed}$ and damping coefficient $c_s = 0.0~\si{\dyne \second \per\centi\meter\cubed}$ ~\cite{FSI_aortic_diss_prestress_Baumler2020-ap}. This boundary condition helps prevent excessive wall motion and nonphysical oscillations \cite{Robin_Moireau2012-sc}. Blood was modeled as a Newtonian fluid with density $\rho=1.06$~g/cm$^3$ and dynamic viscosity $\mu=0.04$~g/(cm\,s). In the FSI simulation, the vessel wall was modeled as a nearly incompressible neo-Hookean material with Elasticity modulus $E=8\times10^6$ ~dyn/cm$^2$, Poisson's ratio 0.49, and density 1.2~g/cm$^3$. No prestress or zero-pressure reference configuration was computed; the image-derived geometry was taken as the stress-free reference configuration. This choice affects wall displacements and may also affect nonlinear convergence and runtime; the performance results apply to this specified model configuration. Both simulations used a time step of $\Delta t=10^{-3}$~s and $\rho_\infty=0.5$. The simulations start from zero velocity and pressure, and the distal time constant of each RCR model, $R_d C\approx1.73$~s, exceeds the cardiac period of $0.968$~s. The pressure field in Fig~\ref{fig:aaaDomain}B, taken after two cardiac cycles, therefore remains below the $90$~mmHg target: several cardiac cycles are required before the outlet pressures become periodic~\cite{RCR_Pfaller2021-oa}. A continuation of the rigid-wall CFD simulation over five cardiac cycles (\nameref{S2_Fig}) confirms this behavior: the outlet flow rates repeat from cycle to cycle to within $1.3\%$ of their peak value by the fifth cycle, whereas the cycle-averaged outlet pressure increases by $25$, $14$, $8.3$, and $4.7$~mmHg between consecutive cycles. This geometric decrease, by a factor of approximately $0.57=\exp(-0.968/1.73)$ per cycle, extrapolates to a periodic mean outlet pressure of approximately $91$~mmHg, consistent with the $90$~mmHg target, and indicates that about eight cycles are needed before the cycle-averaged pressure changes by less than $1$~mmHg between cycles.

In the following sections, we present a performance analysis for an abdominal aortic aneurysm (AAA) model simulated both as a pure CFD problem with rigid walls and as a fully coupled FSI problem using the ALE framework. These two simulations are used to assess total wall-clock time and strong scalability when using different linear solvers and preconditioners available in the FSILS library and in the Trilinos library. Additionally, we present preliminary results comparing the wall-clock times for two Trilinos preconditioners for the AAA CFD case on different architectures.

The CFD setup is identical to the FSI configuration described above, except that a no-slip boundary condition is applied on the vessel wall instead of solving the structural mechanics equations. For both the CFD and FSI cases, we report strong scalability results and wall-clock times obtained with the selected linear solver–preconditioner combinations.
The CPU CFD and FSI studies advance 20 time steps of $10^{-3}$~s from rest, without restarting a previous simulation. MPI process counts range from 4 to 256, with one MPI process per core, on Intel Xeon CPU Max 9480 nodes of Stampede3 (112 cores per node; 1 node up to 64 processes, 2 nodes for 128 and 3 nodes for 256). svMultiPhysics v1.0.0.2 was built from source with the Intel oneAPI 2024.0 compilers, Intel MPI 2021.11, and the Trilinos version pinned by the release (commit \texttt{01c06c6}). Reported times are the total elapsed wall time measured with the Linux \texttt{time} utility, including launch, input and mesh reading, partitioning, initialization, preconditioner setup, time stepping, and output. The time spent outside the time-stepping loop is 18--24~s for CFD and 32--42~s for FSI; for the converged configurations it is below $5\%$ of the total up to 16 (CFD) and 32 (FSI) processes and at most $23\%$ on 256 processes. Because the 20 time steps start from rest, they include the start-up transient of the flow. A continuation of the rigid-wall CFD model over five cardiac cycles (\nameref{S2_Fig}) shows that the cost per cardiac cycle is the same in every cycle to within $5\%$ and that, for this configuration, time steps in developed flow are cheaper than the steps from rest, so the reported times do not underestimate the cost of a periodic simulation.

For all configurations, the Newton iteration cap is 7, with a relative nonlinear tolerance of $10^{-4}$. The Trilinos configurations use GMRES with a relative tolerance of $10^{-4}$, at most 5000 iterations, and a Krylov dimension of 300; the FSILS configurations use the linear-solver settings of the original study, which are included in the input files. The ML preconditioner is the MueLu smoothed-aggregation algebraic multigrid with the default settings of svMultiPhysics: uncoupled aggregation (aggregate size 2--8), at most six levels in a V-cycle, two Gauss--Seidel pre- and post-smoothing sweeps, and a direct (KLU) coarse solver for coarse problems of at most 2000 unknowns. Speedup is computed from the end-to-end time of the same configuration on 4 processes. A time step is counted as converged if Newton's method stops before the iteration cap, or if the final relative residual meets the tolerance. Only runs with all 20 steps converged and no linear solve that reached its iteration cap without meeting its tolerance enter the speedup plots; the wall-time plots also show the remaining runs, marked as incomplete. Elapsed times, convergence counts, and Newton and linear iteration counts for all 98 runs are given in \nameref{S1_Table} and \nameref{S1_Data}, and representative nonlinear residual and linear iteration histories are shown in \nameref{S1_Fig}.

\subsubsection*{AAA: CFD simulation}
\begin{figure}[!ht]
\includegraphics[width=0.95\linewidth]{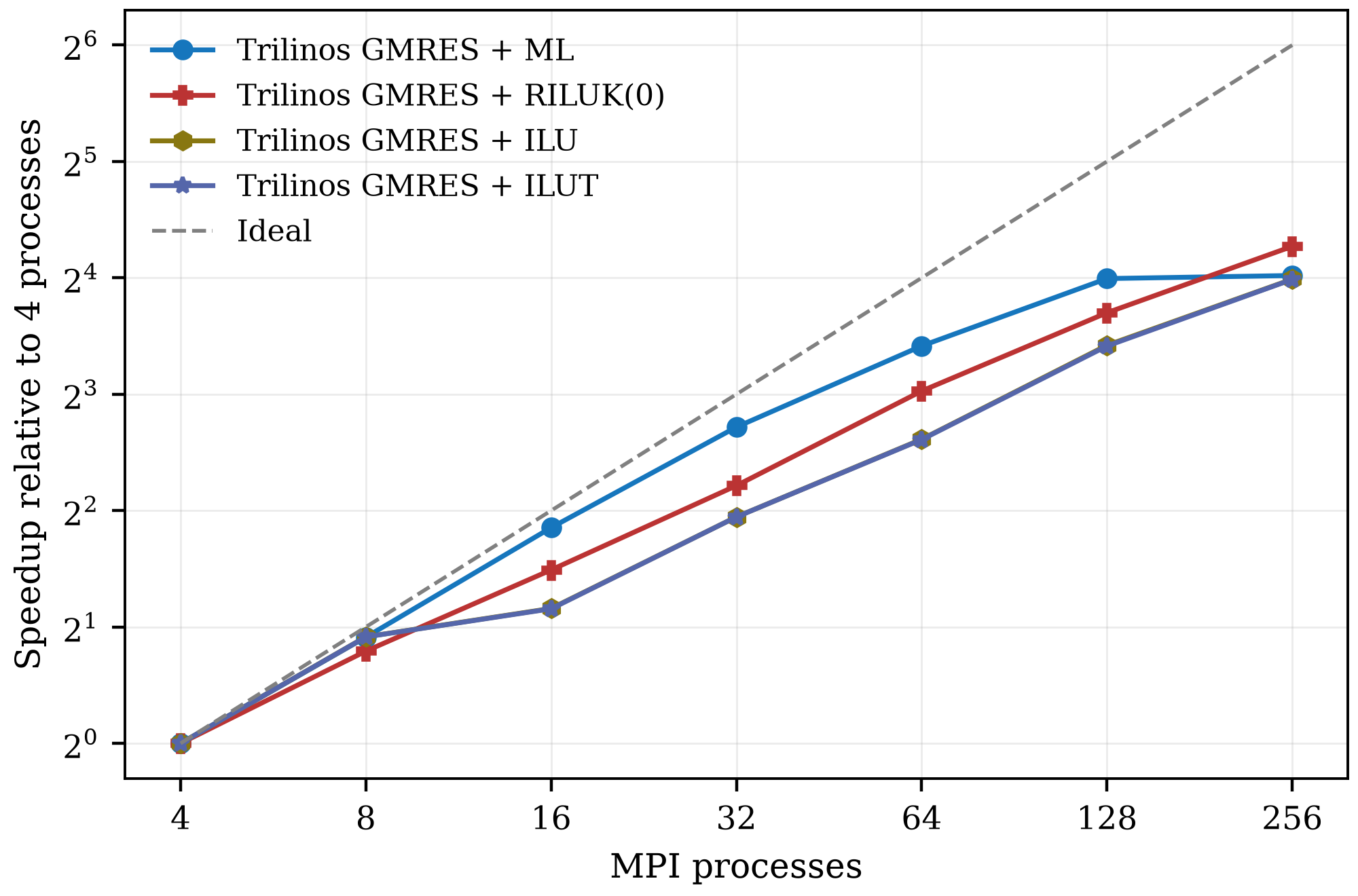}
\caption{{\bf Strong scalability of the AAA CFD simulation.} Speedup of the end-to-end time through 20 steps relative to the same configuration on 4 processes (svMultiPhysics v1.0.0.2). Only configurations with all 20 steps converged and no linear solve reaching its iteration cap are shown. The ideal line is $p/4$.}
\label{fig:scal-cfd}
\end{figure}

Fig~\ref{fig:scal-cfd} shows the scaling of the CFD configurations that satisfy both convergence checks: Trilinos GMRES with ML, RILUK(0), ILU, and ILUT. ML has the lowest end-to-end runtime among them at every process count, decreasing from 1653~s at 4 processes to 104~s at 128 processes. Its runtime does not decrease further at 256 processes (102~s), whereas RILUK(0), ILU, and ILUT, which are two to ten times slower than ML, continue to scale up to 256 processes.

\begin{figure}[!ht]
\includegraphics[width=0.95\linewidth]{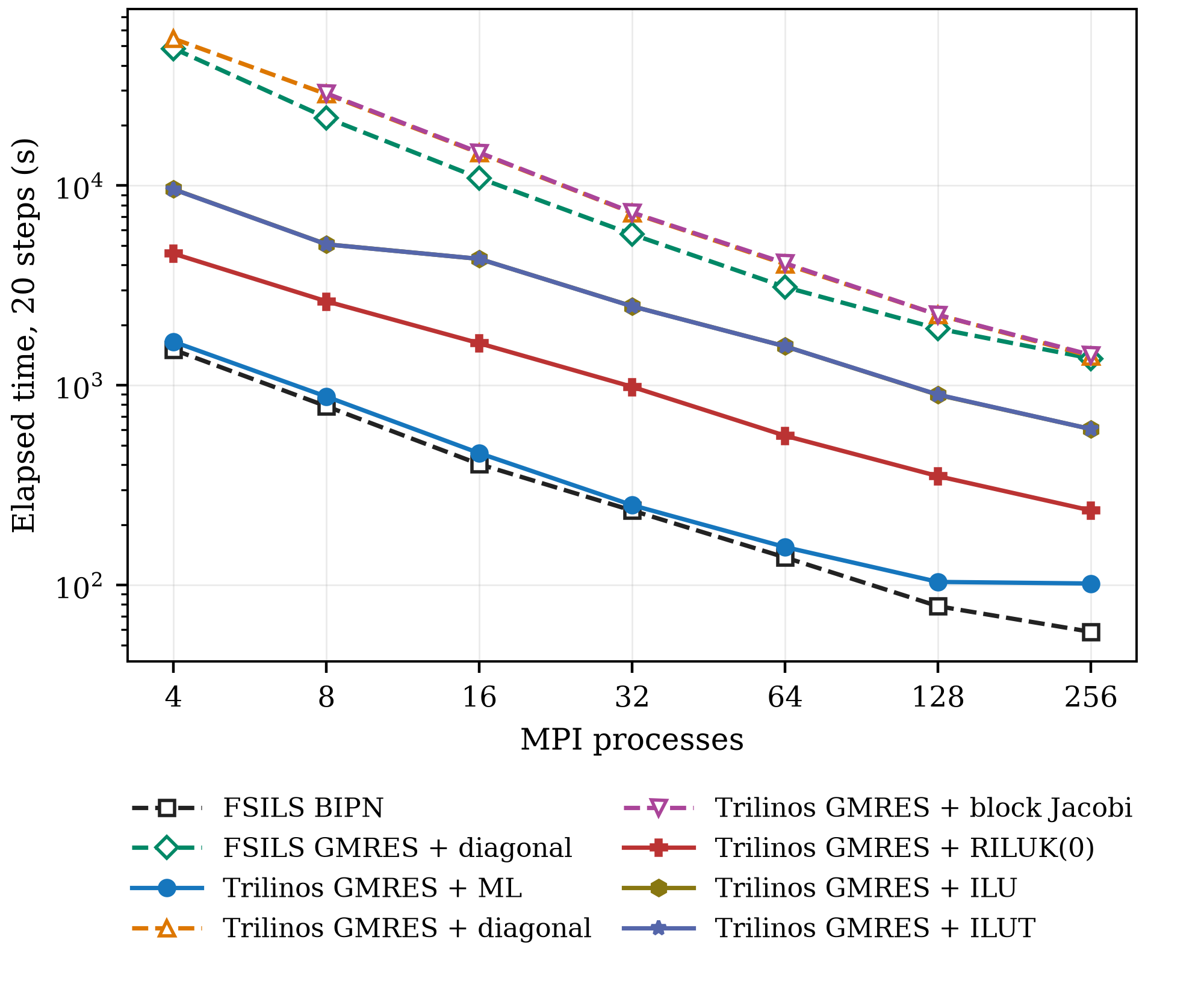}
\caption{{\bf End-to-end elapsed times of the AAA CFD simulation through 20 steps.} Timings (svMultiPhysics v1.0.0.2) include initialization, preconditioner setup, time stepping, and output. Dashed lines with open markers show configurations that did not reach the prescribed tolerances within the iteration limits.}
\label{fig:wall-cfd}
\end{figure}

Fig~\ref{fig:wall-cfd} also shows, as dashed lines, the configurations that did not reach the prescribed tolerances within the iteration limits. This occurs at every process count with nearly the same iteration counts (\nameref{S1_Table}), and therefore reflects the limited effectiveness of these preconditioners for this system rather than the parallel execution. FSILS BIPN is faster than ML at every process count, but only 8--12 of its 20 time steps meet the nonlinear criterion. Its input specifies a loose linear tolerance (0.4) with at most 17 iterations, so each Newton iteration performs only a few linear iterations (\nameref{S1_Fig}); its shorter runtimes therefore do not represent the cost of a converged solution. FSILS GMRES with diagonal preconditioning and the Trilinos diagonal and block-Jacobi configurations meet the nonlinear criterion on 19 of 20 steps but repeatedly reach the linear iteration cap. The ILU and ILUT timings nearly coincide, but similar runtimes do not establish identical effective algorithms. These results replace the conclusion of the original submission that BIPN provides the fastest CFD solution. Previous studies of preconditioned iterative solvers for cardiovascular flows have likewise found that the relative performance of solvers and preconditioners depends strongly on the problem and its boundary conditions~\cite{liu2020,seo2019}.


\subsubsection*{AAA: FSI simulation}
\begin{figure}[!ht]
\includegraphics[width=0.95\linewidth]{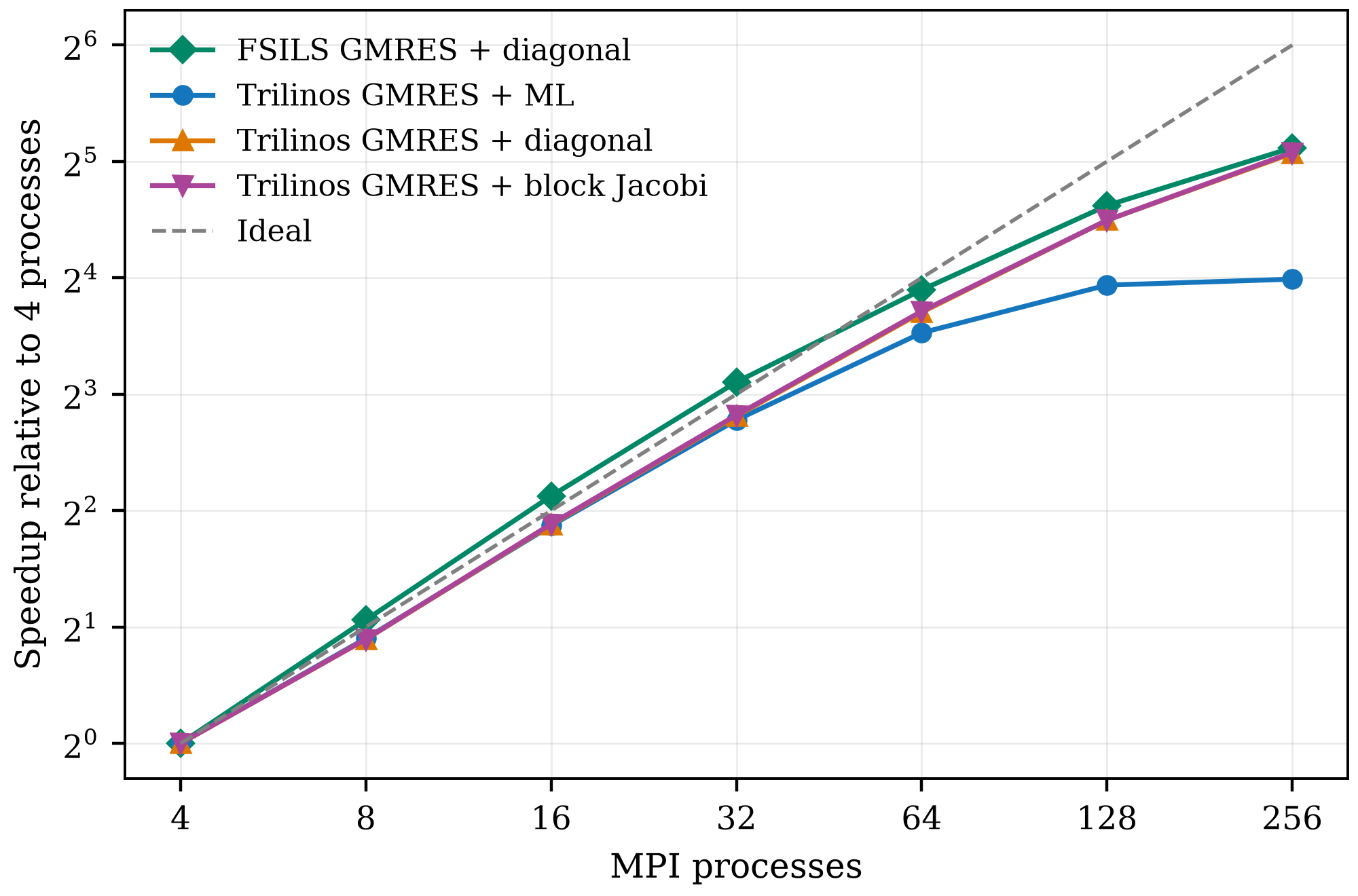}
\caption{{\bf Strong scalability of the AAA FSI simulation.} Speedup of the end-to-end time through 20 steps relative to the same configuration on 4 processes (svMultiPhysics v1.0.0.2). Only configurations with all 20 steps converged and no linear solve reaching its iteration cap are shown. The ideal line is $p/4$.}
\label{fig:scal-fsi}
\end{figure}

Fig~\ref{fig:scal-fsi} shows the scaling of the FSI configurations that satisfy both convergence checks: Trilinos GMRES with ML, diagonal, and block-Jacobi preconditioning, and FSILS GMRES with diagonal preconditioning. ML has the lowest runtime from 4 to 64 processes, decreasing from 3931~s at 4 processes to 341~s at 64 processes, but it scales poorly beyond that point (257~s at 128 and 248~s at 256 processes). FSILS GMRES with diagonal preconditioning scales better and is the fastest configuration at 128 and 256 processes (248~s and 176~s); the Trilinos diagonal and block-Jacobi configurations take 246 and 247~s at 256 processes.

\begin{figure}[!ht]
\includegraphics[width=0.95\linewidth]{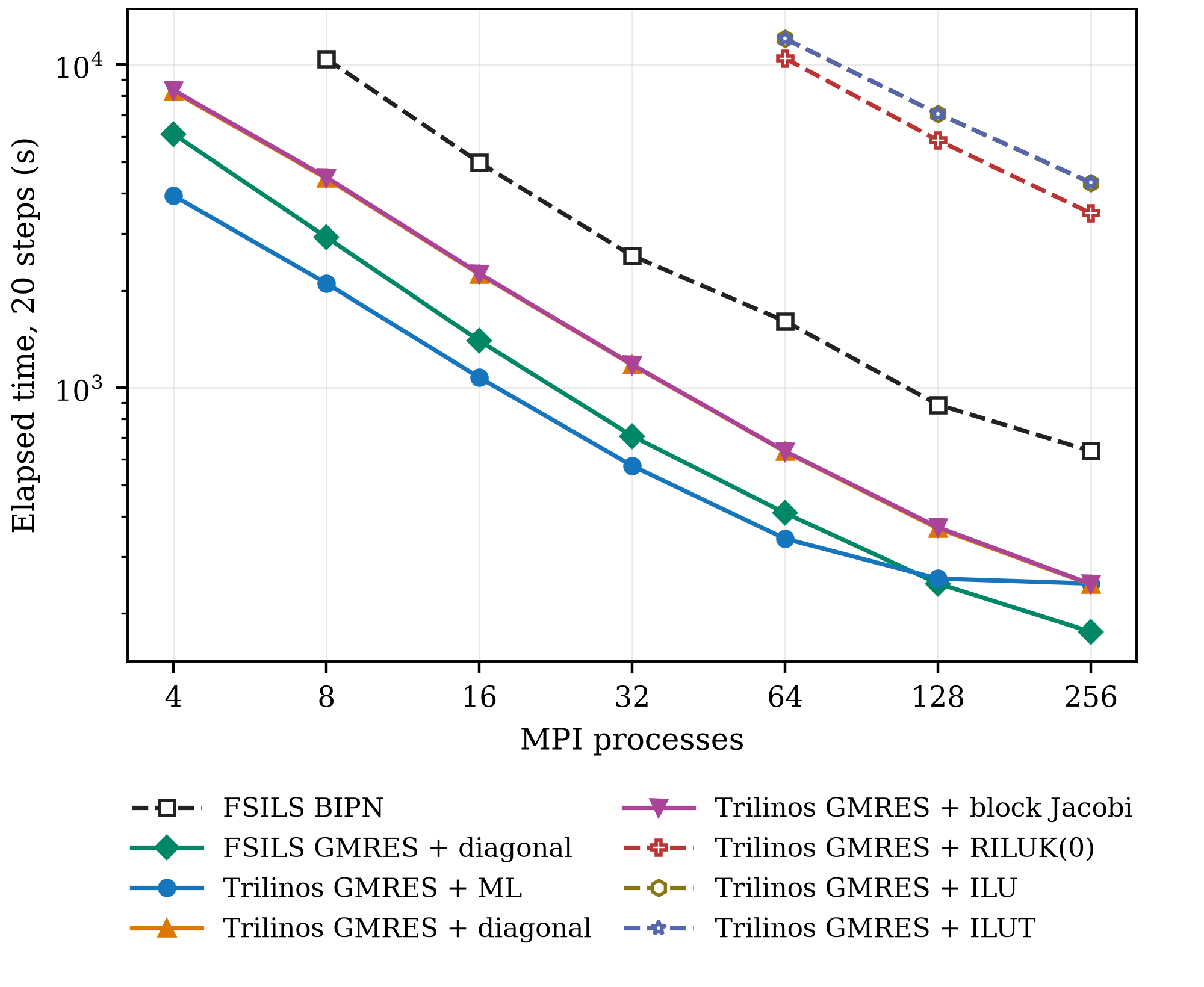}
\caption{{\bf End-to-end elapsed times of the AAA FSI simulation through 20 steps.} Timings (svMultiPhysics v1.0.0.2) include initialization, preconditioner setup, time stepping, and output. Dashed lines with open markers show configurations that did not reach the prescribed tolerances within the iteration limits. BIPN was not run on 4 processes, and ILU, ILUT, and RILUK(0) were run only on 64--256 processes.}
\label{fig:wall-fsi}
\end{figure}

Fig~\ref{fig:wall-fsi} also shows, as dashed lines, the configurations that did not reach the prescribed tolerances; as for CFD, this occurs at every process count tried, with nearly the same iteration counts. BIPN meets the nonlinear criterion at none of the 20 steps on 8--256 processes; the 4-process BIPN run was not performed because its projected runtime exceeded the 48-hour queue limit. The ILU, ILUT, and RILUK(0) runs were performed only at 64--256 processes for the same reason; their fluid--solid linear solves reach the 5000-iteration cap, and none of their 20 steps meets the nonlinear criterion. These runs therefore have no converged 4-process baseline for a speedup.

To separate assembly from the linear solve, we profiled the same inputs with a local development build of svMultiPhysics that records per-stage timers on MPI rank 0; this capability will be made available in svMultiPhysics as a profiling utility (\href{https://github.com/SimVascular/svMultiPhysics/issues/730}{issue \#730}). For ML at 4 processes, assembly, preconditioner setup, and linear solve account for approximately 11\%, 8\%, and 61\% of the elapsed time of the profiled FSI run; at 256 processes the corresponding fractions are 2\%, 15\%, and 61\%. For FSILS GMRES with diagonal preconditioning, the linear solve accounts for approximately 88\% at 4 processes and 72\% at 256 processes. For the CFD case with ML at 4 processes, assembly, preconditioner setup, and linear solve account for approximately 19\%, 16\%, and 57\%. The remainder includes system setup and work outside the profiled stages. These rank-0 measurements characterize this implementation and short workload rather than a synchronized breakdown of work across MPI ranks; the stage fractions for all runs are listed in \nameref{S1_Table}.


\subsubsection*{AAA: Preliminary GPU Performance Comparison} \label{sec:aaa-gpu-performance}
To evaluate the potential benefits of GPU-enabled linear algebra, we performed preliminary wall-clock time comparisons for the AAA CFD case using both CPU-only and GPU-enabled builds of svMultiPhysics. These timings include the simulation setup phase, such as reading input files and mesh data, allocating data structures, and initializing solver objects, in addition to the first two time steps. The short two-step window was chosen to make the lowest-MPI-count cases feasible, since some 4-process CPU runs required on the order of $10^5$ s even for two time steps. One-time initialization contributes a larger fraction of a two-step run than of a long simulation. However, because the CPU and GPU measurements use different platforms, its inclusion does not establish a conservative bound on GPU acceleration. The present measurements do not establish whether solver rankings or speedups persist over longer runs or different phases of the cardiac cycle. Thus, these results should be interpreted as preliminary end-to-end timings rather than fully amortized production-run performance measurements.

\begin{figure}[!ht]
\includegraphics[width=0.9\linewidth]{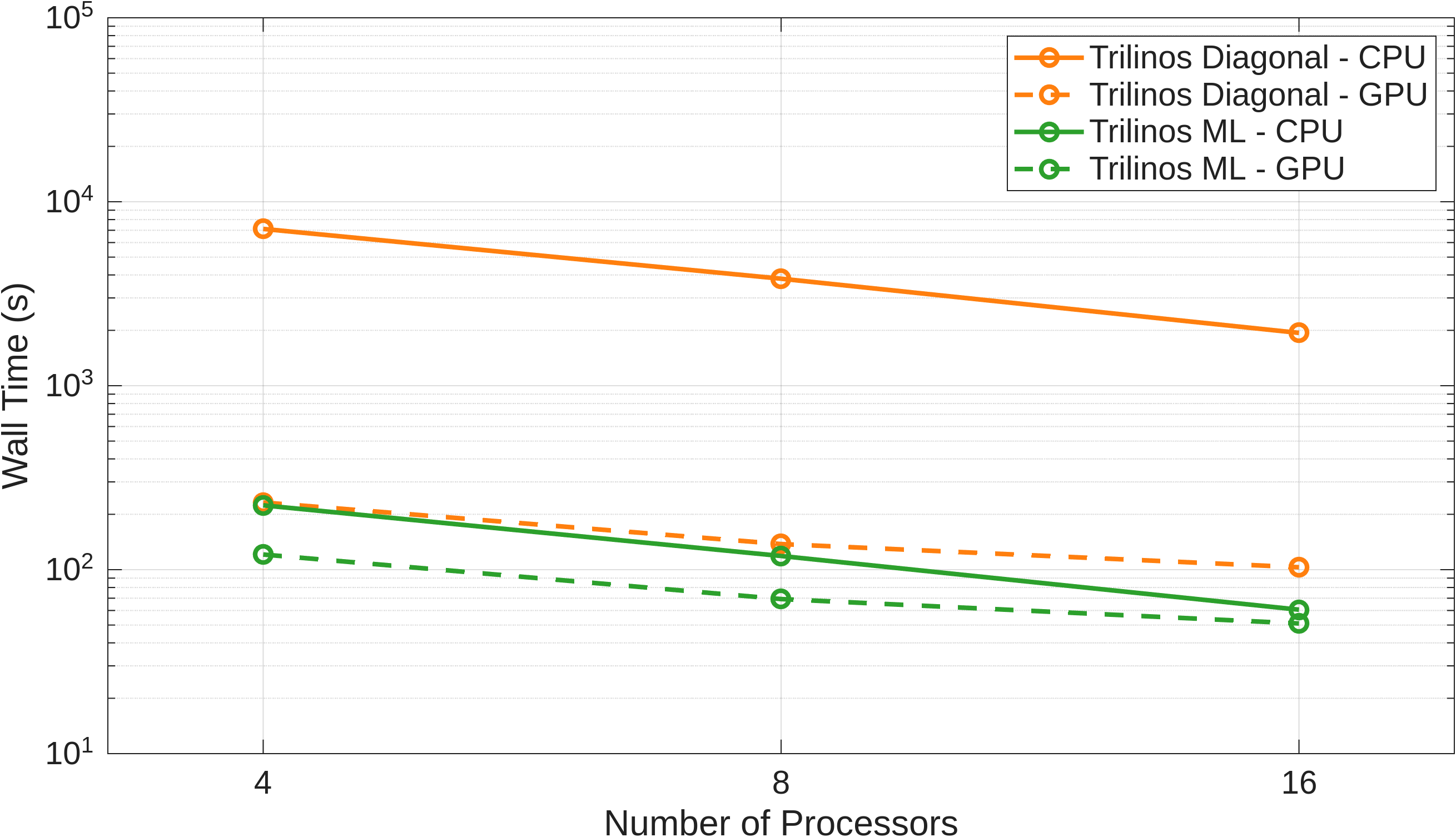}
\caption{{\bf Architecture comparison.}
Wall-clock time comparison between CPU-only and GPU-enabled builds for the AAA CFD simulation using 4, 8 and 16 MPI processes. The CPU runs were performed on Stampede3 Intel Xeon CPU Max 9480 nodes with one core per MPI process, while the GPU-enabled runs were performed on Sherlock nodes equipped with NVIDIA Tesla A40 GPUs, one GPU per MPI process. Times are reported after two steps, using an earlier development version of svMultiPhysics. Because the GPU-accelerated part of the computation is the external linear solver and preconditioner, we expect qualitatively similar behavior with the current release. Linear solver and preconditioner operations are executed on GPU, while finite element assembly remains on the CPU.}
\label{fig:wall-gpu} 
\end{figure}

The CPU simulations were performed on Stampede3 Intel Xeon CPU Max 9480 nodes, whereas the GPU-enabled runs were performed on Sherlock nodes equipped with NVIDIA Tesla A40 GPUs. Two platforms were used because our CPU allocation was on Stampede3, whereas A40 GPUs were available to us only on Sherlock; CPU-only runs on the Sherlock host processors were not performed. The two platforms therefore do not constitute an otherwise identical software and hardware environment; instead, the comparison is intended to assess representative CPU-only and GPU-enabled execution on available HPC resources. The CPU runs used one core per MPI process, whereas the GPU-enabled runs used one NVIDIA Tesla A40 GPU per MPI process, corresponding to 4, 8, and 16 GPUs for the three cases shown. In both cases, identical MPI configurations and solver settings were used. Equal MPI process counts do not represent equal hardware resources or cost in this CPU--GPU comparison. For this preliminary study, we report results using two Trilinos preconditioners - Diagonal and ML. \figurename~\ref{fig:wall-gpu} shows the total runtime after completion of two time steps for 4, 8 and 16 MPI processes.

A consistent reduction in wall-clock time is observed when using GPUs across all processor counts for both preconditioners. The performance improvement is particularly pronounced for the diagonal preconditioner, which decreases from approximately $7{,}000$ s to $230$ s on 4 MPI processes, from $3{,}800$ s to $130$ s on 8 processes, and from $1{,}900$ s to $100$ s on 16 processes, corresponding to roughly $19$--$30\times$ speedups. The ML preconditioner, which is already more efficient on CPUs, also benefits from GPU acceleration, with wall-clock times reduced from approximately $220$ s to $120$ s, $115$ s to $70$ s, and $60$ s to $50$ s over the same processor counts, corresponding to approximately $1.2$--$1.8\times$ speedups. Because the diagonal preconditioner is the slowest CPU configuration, the large ratios obtained with it overstate the practical benefit of GPU execution; the relevant comparison is between the fastest CPU and GPU-enabled configurations, both ML, which differ by approximately $1.2$--$1.8\times$. Overall, the GPU-enabled configurations have lower measured runtimes in this comparison, but hardware and software differences between the platforms prevent attribution of the full reduction to GPU offloading alone. The GPU results reported here are based on NVIDIA A40 hardware.

\subsection*{Cardiac electrophysiology model description}

To further demonstrate the versatility of svMultiPhysics, we consider a cardiac electrophysiology simulation on a patient-specific biventricular geometry with congenital heart disease (CHD). The three-dimensional ventricular anatomy was reconstructed from computed tomography (CT) imaging. Myocardial fiber architecture was generated using the rule-based approach of Doste et al.~\cite{doste2019}, which provides fiber orientations in biventricular geometries with outflow tracts. 

In the myocardium, electrical conduction is modeled using $D_{\text{iso}} = 0.05$~mm$^2$/ms and $D_{\text{f}} = 0.1$~mm$^2$/ms ($D_{\text{s}}=D_{\text{n}}=0$). 
In addition to the three-dimensional myocardial domain, a one-dimensional Purkinje network was generated using the method proposed by Sahli Costabal et al.~\cite{sahli2016}. For the Purkinje network, conduction is modeled using a constant conductivity $D_{\text{purk}} = 3.0$~mm$^2$/ms, reflecting the fast-conducting nature of Purkinje fibers. This fast-conducting network was embedded on the endocardial surfaces and coupled to the ventricular myocardium through Purkinje--myocardial junctions. The combined 1D--3D electrophysiology model allows rapid activation through the Purkinje system followed by propagation into the ventricular tissue, providing a physiologically relevant activation sequence for ventricular depolarization. 

We modeled electrical activation by solving the monodomain equation using an operator-splitting strategy with a time step of $0.1$ ms. Cellular membrane dynamics in the myocardium were represented using the ten Tusscher--Panfilov ionic model~\cite{tentusscher2004,tentusscher2006}, with epicardial cell parameters prescribed. In the Purkinje network, we used ten Tusscher--Panfilov models with endocardial parameters, as a dedicated Purkinje cell model is not yet available (see Current limitations). Excitation was initiated by applying a current stimulus to the roots of the left and right ventricular Purkinje fractal tree networks. The action potential then propagates rapidly through the Purkinje network before entering the myocardium through the Purkinje--myocardial junctions, where it triggers ventricular activation. 

The biventricular mesh consists of 320,200 linear tetrahedral elements and 77,290 nodes, with a mean edge length of $1.49$~mm (5th--95th percentile $0.82$--$2.29$~mm); the Purkinje network consists of 4194 line elements with a mean length of $0.79$~mm and 308 Purkinje--myocardial junctions. Given the relatively coarse mesh, we considered non-Gaussian quadrature rules, following the approach introduced in Tikeno{\u{g}}ullar{\i} et al.~\cite{tikenogullari2024}. Linear systems were solved with GMRES and CG to a relative tolerance of $10^{-12}$. This example is used to assess computational performance, and its activation times should not be interpreted as mesh-converged.

\begin{figure}[!ht]
\includegraphics[width=0.9\linewidth]{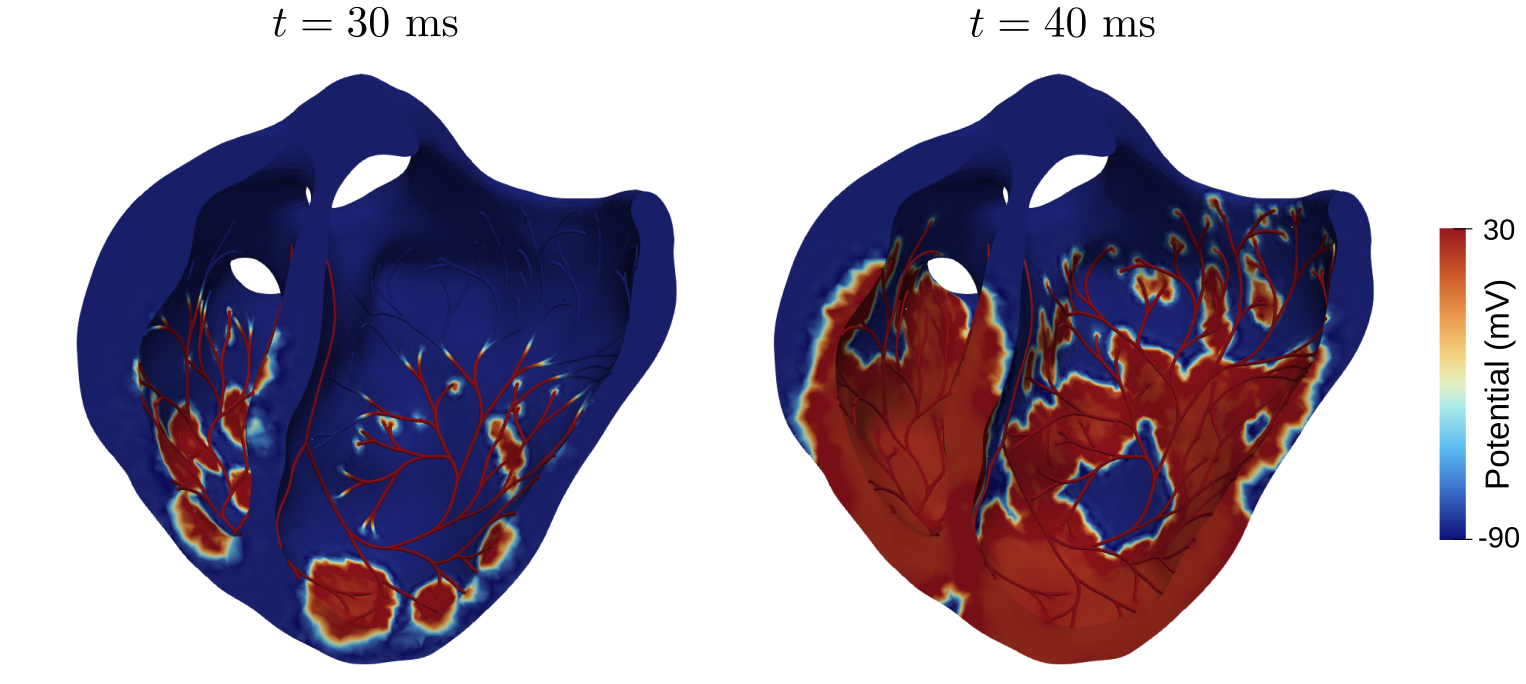}
\caption{{\bf Cardiac electrophysiology model.}
Patient-specific biventricular electrophysiology simulation. Two representative time points are shown to illustrate depolarization of the Purkinje network and the subsequent propagation of electrical activity through the myocardium.}
\label{fig:ep-domain}
\end{figure}

Fig~\ref{fig:ep-domain} illustrates the patient-specific electrophysiology model. The myocardial domain is shown together with the generated Purkinje network distributed inside the ventricles. The two views highlight the anatomical complexity of the biventricular geometry and the branching structure of the fast-conducting network used to drive activation.

\subsubsection*{Cardiac electrophysiology simulation}
Next, we evaluate the computational performance of svMultiPhysics for this cardiac electrophysiology case. In contrast to the AAA CFD and FSI examples, this problem involves the solution of a reaction--diffusion type system in a complex biventricular geometry coupled to a distributed Purkinje network. This example demonstrates that svMultiPhysics is not limited to fluid and solid mechanics applications, but also supports large-scale simulations of cardiac electrical activation within the same computational framework.

For the performance study, we report the total wall-clock time after completion of 500 time steps, corresponding to 50 ms of simulated time with a time step size of $0.1$ ms, computed with svMultiPhysics v1.0.0.2. Strong scalability is assessed by doubling the number of MPI processes starting from 4 processors, and the speedup is normalized by the wall-clock time measured for 500 steps on 4 processors. All simulations were performed using the same electrophysiology model, fiber architecture, Purkinje network, and stimulation protocol described above.

\begin{figure}[!ht]
\includegraphics[width=0.95\linewidth]{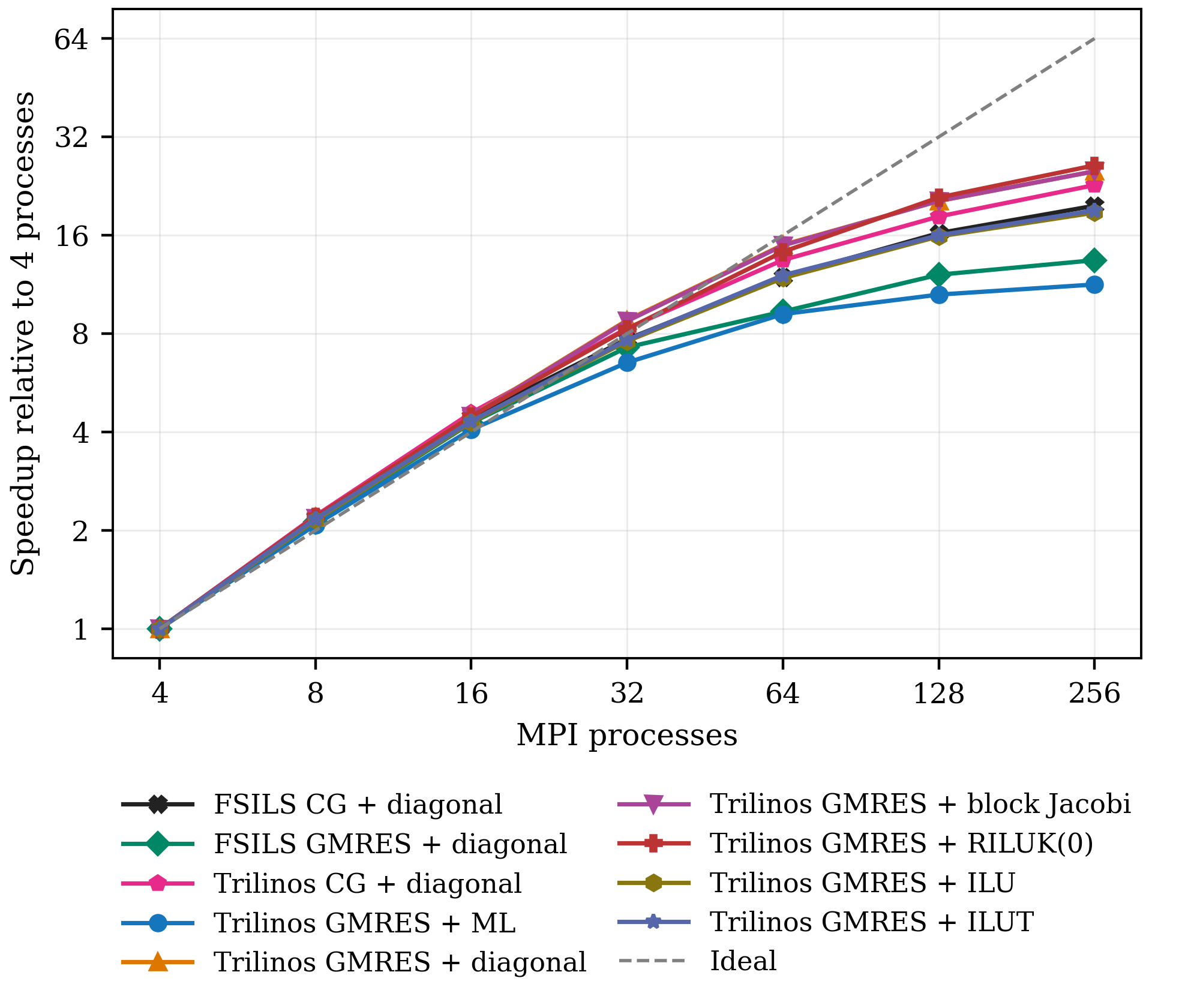}
\caption{{\bf Strong scalability for EP simulation.}
Strong scalability results for the cardiac electrophysiology simulation. The reported speedup is computed using the wall-clock time after 500 time steps and normalized by the corresponding wall-clock time obtained with 4 processors.}
\label{fig:scal-ep}
\end{figure}

\begin{figure}[!hb]
\includegraphics[width=0.95\linewidth]{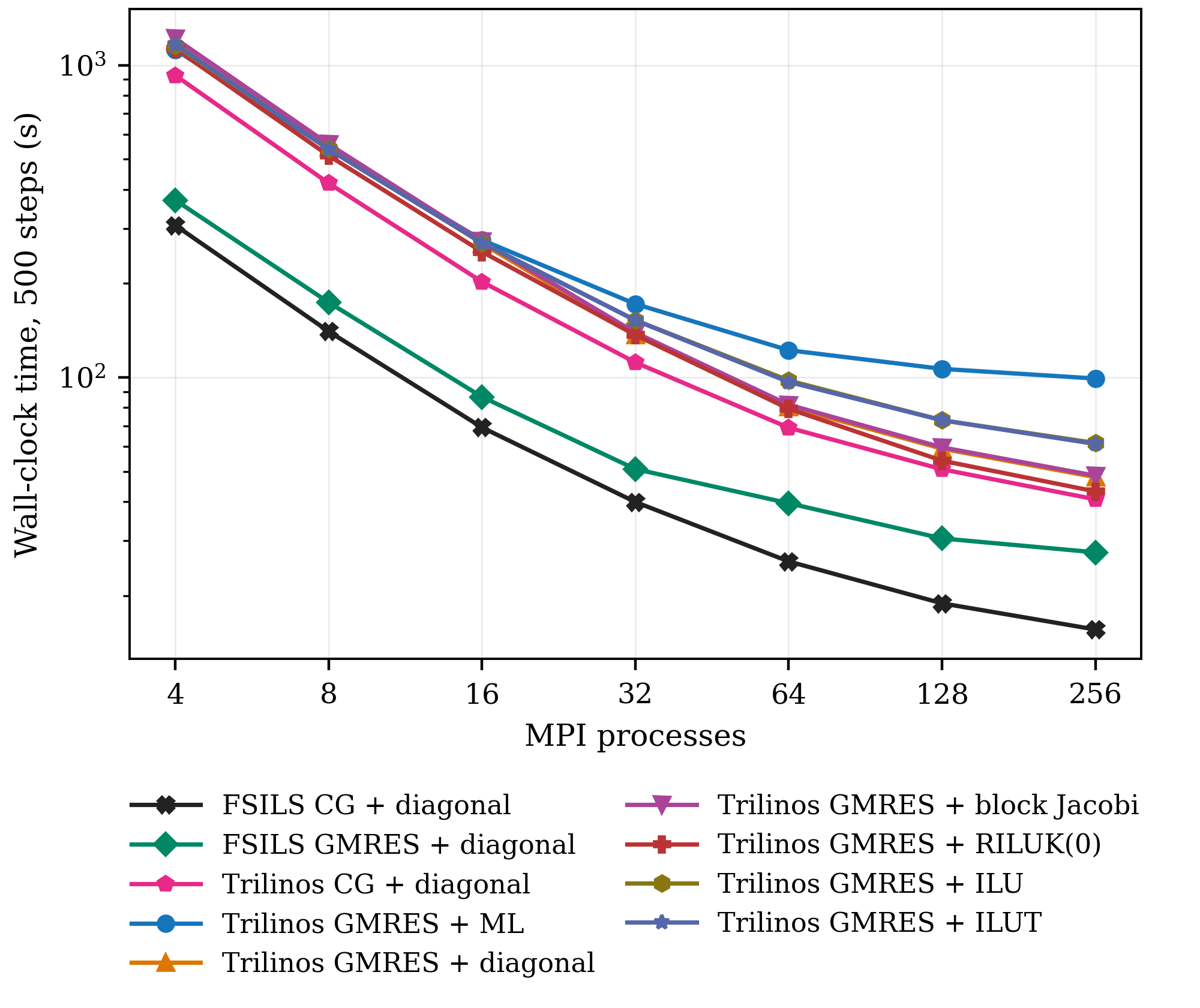}
\caption{{\bf Wall-clock times for EP simulation.}
Wall-clock times for the cardiac electrophysiology simulation reported after completion of 500 time steps, corresponding to 50 ms of simulated time.}
\label{fig:wall-ep}
\end{figure}

\figurename~\ref{fig:scal-ep} shows the strong scalability behavior for the electrophysiology simulation. The results indicate how efficiently the solver handles the repeated solution of the diffusion problem together with the ionic update over hundreds of time steps in an anatomically realistic biventricular model. \figurename~\ref{fig:wall-ep} presents the corresponding wall-clock times after 500 time steps. Across all processor counts, the FSILS CG solver with diagonal preconditioning yields the lowest wall-clock time (307~s on 4 processes and 16~s on 256 processes), followed by FSILS GMRES with diagonal preconditioning. Among the Trilinos-based approaches, CG with diagonal preconditioning is the fastest, GMRES with the diagonal, block-Jacobi, and RILUK(0) preconditioners performs similarly, while the multilevel (ML) preconditioner incurs the highest cost from 16 processes upward (99~s on 256 processes). For identical solver--preconditioner combinations, the FSILS implementations are faster than their Trilinos counterparts, by a factor of 1.7--3.3 for GMRES and 2.6--3.0 for CG with diagonal preconditioning. The time per linear iteration is similar in the two libraries (about 8.5~ms for GMRES on 4 processes); the difference comes from the number of iterations: with the same tolerance of $10^{-12}$, the FSILS solvers needed about 43,000 linear iterations over the 500 time steps, compared with 143,000--145,000 for the Trilinos solvers with diagonal preconditioning. The speedup from 4 to 256 processes ranges from 11 (Trilinos GMRES with ML) to 26 (Trilinos GMRES with RILUK(0)); the FSILS solvers, which are the fastest on few processes, reach 13 (GMRES) and 20 (CG). Overall, these results highlight the efficiency of the lightweight FSILS solvers for this class of problems, while the Trilinos-based implementations offer a broader choice of solvers and preconditioners.

Together, these results provide a representative measure of the computational cost of cardiac electrophysiology simulations in svMultiPhysics and complement the CFD and FSI performance studies presented earlier. The results also confirm the ability of svMultiPhysics to handle anatomically informed conductivity tensors, physiologically motivated Purkinje activation pathways, and realistic ventricular-scale electrophysiology simulations within a unified finite element solver framework.\newline

\subsection*{Current limitations}
Several limitations of the present version should be noted. First, the electrophysiology module solves the monodomain equation only; bidomain formulations and an explicit extracellular potential field are not available, and electrocardiograms are computed with the pseudo-ECG approximation of Eq~(\ref{eq:ecg}). Second, four ionic models are currently implemented (Table~\ref{table:electrophysiology}): the biophysically detailed ten Tusscher--Panfilov ventricular model (with endocardial, midmyocardial, and epicardial parameter sets) and three phenomenological models. Dedicated atrial, Purkinje, and sinoatrial cell models are not yet available, so ventricular models are also used in the conduction system. Ionic models cannot be imported from CellML and must be implemented in C++ through the \texttt{IonicModel} class interface. Third, in the resistive immersed surface approach, valve leaflet motion is prescribed kinematically rather than computed from leaflet mechanics during the coupled simulation. Fourth, GPU execution is currently limited to the Trilinos linear solvers and preconditioners; finite element assembly is performed on CPUs. Fifth, porous-media flow is restricted to rigid porous media. Finally, much of the code retains the procedural structure inherited from svFSI. Bidomain and CellML support, atrial and Purkinje cell models, and GPU-resident assembly are priorities of ongoing development.

\section*{Conclusions}
In this work, we presented svMultiPhysics, an open-source, MPI-parallel finite element solver designed for cardiovascular applications and built to support tightly coupled multiphysics simulations. The solver is implemented in C++ and provides a unified computational framework in which the governing equations of fluid dynamics, solid mechanics, scalar transport, and cardiac electrophysiology can be solved either independently or in strongly coupled configurations. This integrated design represents one of the core strengths of svMultiPhysics, enabling researchers to investigate complex physiological interactions within a consistent numerical and data-structural environment. The present implementation was verified against published benchmarks for incompressible flow, fluid--structure interaction, cardiac electrophysiology, and cardiac mechanics, including direct comparisons with results of its Fortran predecessor (Appendices~\ref{app:cfd-benchmark}--\ref{app:cardiac-mechanics-benchmark}). These results complement the broader validation history of the SimVascular solver ecosystem and its precursor implementations, which has demonstrated agreement with experimental and in vivo measurements in patient-specific coronary aneurysm hemodynamics in Kawasaki disease, vascular fluid–structure interaction against in vitro 4D-flow MRI, pressure-drop estimation in aortic coarctation, and heart-valve hemodynamics against in vitro 4D-flow MRI data\cite{kung2014validation,lan2023rucvalidation,nair2024coa_validation,seresti2025cta}.

A key component of the solver is its flexible linear algebra interface. svMultiPhysics incorporates its own linear solver library (FSILS) and provides seamless interoperability with two major scientific libraries, PETSc and Trilinos. This approach offers users a broad spectrum of linear solvers and preconditioners, allowing the selection of methods best suited to the problem being solved. Such flexibility is fundamental for tackling the large, non-linear, coupled, and often ill-conditioned systems that arise in cardiovascular simulations.

To illustrate the solver’s capabilities, we considered three representative applications, a CFD and an FSI simulation of a patient-specific abdominal aortic aneurysm; and a cardiac electrophysiology simulation on a patient-specific biventricular geometry with CHD. These examples highlight the ability of svMultiPhysics to handle realistic geometries, non-linear multiphysics coupling, and large-scale parallel execution. The 20-step CPU studies with svMultiPhysics v1.0.0.2 show good strong scaling at lower process counts and saturation of the ML preconditioner at higher counts. ML is the fastest converged CFD configuration at all tested process counts and the fastest converged FSI configuration up to 64 processes, while FSILS GMRES with diagonal preconditioning is fastest for FSI at 128 and 256 processes. These results apply to the specified short workload; a five-cycle CFD simulation shows that the cost per cardiac cycle is established in the first cycle, and that the 20 steps from rest are more expensive per step than the same phase of developed flow (\nameref{S2_Fig}). The wall-clock time comparison across different linear solvers and preconditioners further illustrates the flexibility and efficiency of the available solution strategies.

The preliminary CPU--GPU comparison reports end-to-end runtimes, including setup and two time steps, on different computing platforms. GPU-enabled configurations are faster for the selected Trilinos preconditioners in this workload, but these results do not isolate linear-solver acceleration or establish a production-run speedup over the fastest CPU solver. Finite element assembly remains on the CPU.

The current C++ implementation of svMultiPhysics originates from a systematic translation of the previous Fortran-based solver svFSI~\cite{svfsi}. Appendix~\ref{app:fsi-benchmark} compares the beam-tip oscillations computed by svMultiPhysics for a two-dimensional FSI benchmark with published results from the predecessor Fortran implementation and an isogeometric formulation. Appendix~\ref{app:whole-heart-coupling} additionally demonstrates active myocardial contraction coupled to whole-heart blood flow, immersed valve surfaces, and a closed-loop circulation. Appendix~\ref{app:electromechanics} demonstrates electromechanical coupling, in which the active tension is computed from the simulated calcium transient rather than prescribed. Ongoing development efforts are focused on achieving a fully object-oriented design to improve modularity, maintainability, and extensibility. This refactoring will ensure that new physics modules, numerical strategies, and solver components can be incorporated with minimal friction, supporting the long-term sustainability and evolution of the project.

As the multiphysics backend of the SimVascular project, svMultiPhysics provides the foundation for advanced cardiovascular modeling workflows. By documenting its governing equations, numerical formulations, code architecture, and representative applications, this paper aims to serve both users and developers. Our goal is to provide a clear and comprehensive reference that supports reproducibility, encourages community contributions, and fosters the continued growth of an open-source platform for cardiovascular simulation research.

svMultiPhysics is actively maintained, extensively documented, and openly developed on GitHub, ensuring transparency and long-term sustainability. Its unified multiphysics capabilities, coupled with robust parallel performance, position it as a versatile tool for investigating complex cardiovascular problems—from fundamental biomechanical studies to clinical applications.

\clearpage
\begingroup
\newcounter{benchmarksection}
\renewcommand{\thebenchmarksection}{\Alph{benchmarksection}}
\makeatletter
\@addtoreset{figure}{benchmarksection}
\@addtoreset{table}{benchmarksection}
\makeatother
\renewcommand{\thefigure}{\thebenchmarksection\arabic{figure}}
\renewcommand{\thetable}{\thebenchmarksection\arabic{table}}
\ifdefined\theHfigure
  \renewcommand{\theHfigure}{benchmark.\thebenchmarksection.\arabic{figure}}
\fi
\ifdefined\theHtable
  \renewcommand{\theHtable}{benchmark.\thebenchmarksection.\arabic{table}}
\fi

\refstepcounter{benchmarksection}\label{app:cfd-benchmark}
\section*{Appendix \thebenchmarksection. Two-Dimensional Flow Past a Circular Cylinder}

\subsection*{Problem definition}
We verify the incompressible flow solver using two-dimensional laminar flow past a circular cylinder at a Reynolds number of $\mathrm{Re}=100$, a standard test for unsteady incompressible flow solvers. At this Reynolds number the wake develops periodic von K\'arm\'an vortex shedding, and the mean drag coefficient, root-mean-square (RMS) lift coefficient, and Strouhal number are well established~\cite{tumkur2013cylinder,kamensky2015immersogeometric,tsetoglou2025volume,deng2019cylinder,qu2013cylinder}.

A cylinder of diameter $D=1$ is centered at the origin of the rectangular domain $[-15,25]\times[-12,12]$~(Fig~\ref{fig:cylinder-domain}), so the inlet lies $15D$ upstream of the cylinder center, the outlet $25D$ downstream, and each lateral boundary $12D$ from the center (blockage ratio $1/24$). All quantities are nondimensional, with density $\rho=1$, dynamic viscosity $\mu=0.01$, and freestream velocity $U_\infty=1$, giving $\mathrm{Re}=\rho U_\infty D/\mu=100$. A uniform velocity $(U_\infty,0)$ is prescribed at the inlet. The outlet has a zero-traction condition with backflow stabilization (coefficient $0.2$). The upper and lower boundaries are slip walls, with zero normal velocity and unconstrained tangential velocity, and a no-slip condition is imposed on the cylinder surface.

\begin{figure}[!htbp]
  \centering
  \includegraphics[width=\linewidth]{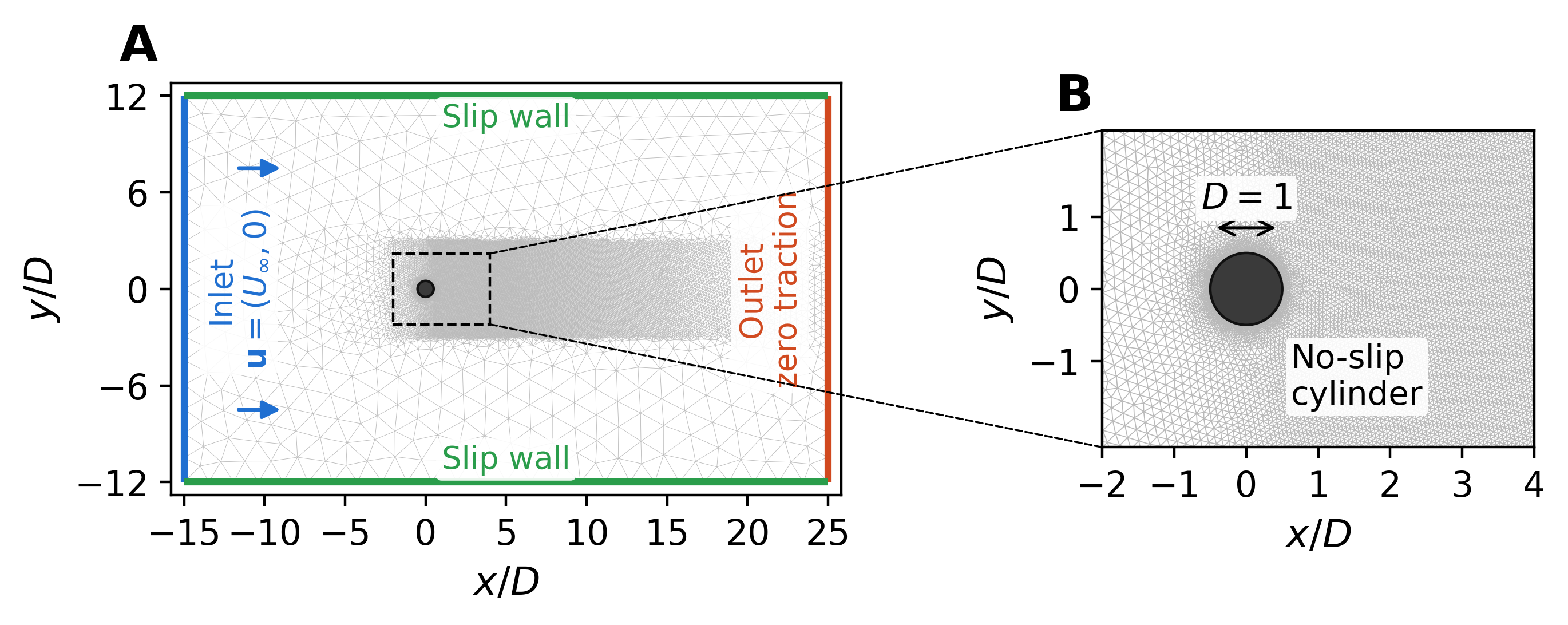}
  \caption{\textbf{Computational domain for flow past a circular cylinder.} (A) Rectangular domain $[-15,25]\times[-12,12]$ (in units of the cylinder diameter $D$) with its boundary conditions, shown with the coarse mesh. (B) Close-up of the cylinder ($D=1$, centered at the origin) and the mesh refinement around it and in the wake.}
  \label{fig:cylinder-domain}
\end{figure}

\subsection*{Discretization and solution procedure}
The domain was discretized with three unstructured meshes of linear triangles generated with a Frontal--Delaunay algorithm and graded toward the cylinder and wake. The prescribed element size on the cylinder wall decreases by a factor of $1.5$ between successive meshes ($0.012375D$, $0.00825D$, and $0.0055D$), and the element count grows by a factor of approximately $2.25$ (Table~\ref{tab:cylinder}). The flow was solved with the stabilized formulation described in the main text, using the generalized-$\alpha$ method with $\rho_\infty=0.5$ and a time step of $\Delta t=0.005$ for $30{,}000$ steps ($tU_\infty/D=150$). The nonlinear relative residual tolerance was $10^{-5}$, with a minimum of $2$ and a maximum of $8$ Newton iterations per time step, and the linear systems were solved with the BIPN algorithm~\cite{moghadam2015}.  All simulations were performed with svMultiPhysics v1.0.0.2.

The drag and lift coefficients, $C_D=2F_x/(\rho U_\infty^2 D)$ and $C_L=2F_y/(\rho U_\infty^2 D)$, were obtained by integrating the pressure and viscous tractions over the cylinder surface. Statistics were computed over the last three complete shedding cycles on each mesh: the time-averaged drag $\overline{C}_D$, the RMS lift $C_{L,\mathrm{rms}}$ (after removing the mean), and the Strouhal number $\mathrm{St}=D/(U_\infty T)$, where $T$ is the mean shedding period. The time-averaged surface pressure coefficient, $\overline{C}_p=(\overline{p}_w-\overline{p}_\infty)/(\tfrac{1}{2}\rho U_\infty^2)$, uses the mean inlet pressure as the reference and is reported as a function of the angle $\theta$ measured from the upstream stagnation point, averaged over the upper and lower surfaces.

\begin{table}[!htbp]
  \centering
  \caption{\textbf{Meshes and integrated quantities for flow past a cylinder at $\mathrm{Re}=100$.}
  Published values are the range reported in~\cite{tumkur2013cylinder,kamensky2015immersogeometric,tsetoglou2025volume,deng2019cylinder,qu2013cylinder}.}
  \label{tab:cylinder}
  \begin{tabular}{lrrccc}
    \hline
    Mesh & Nodes & Elements & $\overline{C}_D$ & $C_{L,\mathrm{rms}}$ & St \\
    \hline
    Coarse & 23,837 & 47,327 & 1.372 & 0.230 & 0.167 \\
    Medium & 54,050 & 107,582 & 1.362 & 0.230 & 0.168 \\
    Fine & 120,353 & 239,933 & 1.362 & 0.235 & 0.169 \\
    \hline
    Published & & & 1.318--1.381 & 0.222--0.237 & 0.164--0.171 \\
    \hline
  \end{tabular}
\end{table}

\subsection*{Results}
Fig~\ref{fig:cylinder} shows the force histories and the time-averaged surface pressure on the three meshes. After the initial transient, vortex shedding develops between $tU_\infty/D\approx40$ and $70$ and becomes periodic by $tU_\infty/D\approx100$. The integrated quantities show limited changes over the tested refinements: $\overline{C}_D$ changes by $0.71\%$ from the coarse to the medium mesh and by $0.05\%$ from the medium to the fine mesh, and the mean shedding period changes by $0.83\%$ and $0.28\%$ over the same refinements, comparable to the $0.05$ interval between saved solutions. The $\overline{C}_p$ distributions nearly coincide, with a minimum of approximately $-1.20$ near $\theta=84^\circ$ and a base pressure of approximately $-0.76$ on the fine mesh. The largest difference between meshes ($0.044$, coarse versus fine) occurs at the stagnation point.

The resulting $\overline{C}_D$, $C_{L,\mathrm{rms}}$, and $\mathrm{St}$ are within the published ranges (Table~\ref{tab:cylinder}) while the differences between literature values may reflect the lateral confinement of the present domain (blockage ratio $1/24$), as well as temporal discretization and other differences in numerical setup; the present mesh study does not isolate these contributions. Sensitivity to the domain size and time step was not assessed here.

\begin{figure}[!htbp]
  \centering
  \includegraphics[width=\linewidth]{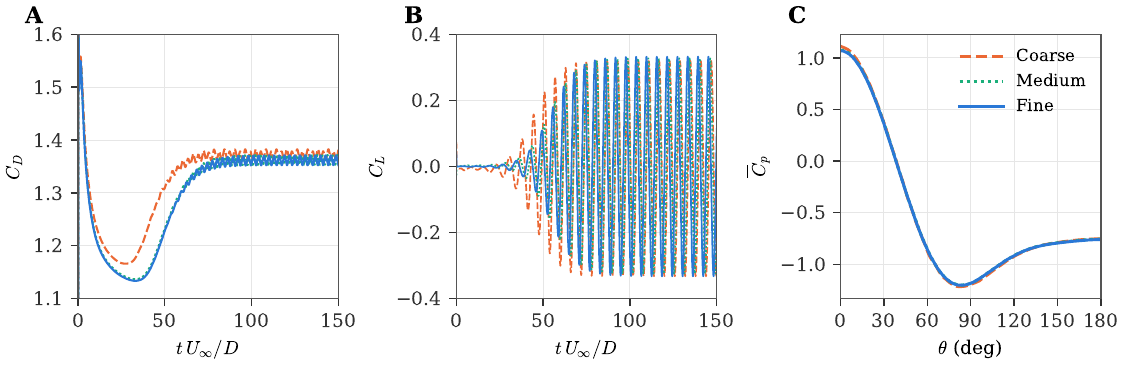}
  \caption{\textbf{Flow past a circular cylinder at $\mathrm{Re}=100$ on three meshes.}
  (A) Drag coefficient and (B) lift coefficient versus nondimensional time. The impulsive-start value of $C_D$ at the first saved step lies outside the plotted range. (C) Time-averaged surface pressure coefficient over the last three shedding cycles versus the angle $\theta$ from the upstream stagnation point, averaged over the upper and lower surfaces.}
  \label{fig:cylinder}
\end{figure}

\clearpage
\refstepcounter{benchmarksection}\label{app:fsi-benchmark}
\section*{Appendix \thebenchmarksection. FSI Benchmark}

\subsection*{Problem definition}
We assess the fluid--structure interaction (FSI) implementation in svMultiPhysics using the two-dimensional benchmark of flow past a flexible beam attached to a fixed square block. This benchmark was introduced by Wall~\cite{wall1999fsi} and subsequently studied by Bazilevs et al.~\cite{bazilevs2008fsi} and Esmaily-Moghadam et al.~\cite{moghadam2015}. Vortex shedding behind the block drives large transverse oscillations of the beam. Comparison with the published results of Esmaily-Moghadam et al. provides a check against the predecessor Fortran implementation, while the isogeometric results of Bazilevs et al. provide an additional numerical reference.

The fluid domain is a rectangular channel of length $19.5\,\mathrm{cm}$ and height $12\,\mathrm{cm}$~(Fig~\ref{fig:fsi-domain}). A fixed square block of side length $D=1\,\mathrm{cm}$ is centered across the channel, with its upstream face $4.5\,\mathrm{cm}$ downstream of the inlet. An initially straight beam of length $4\,\mathrm{cm}$ and thickness $0.06\,\mathrm{cm}$ extends downstream from the center of the block's rear face. The beam is clamped at its attachment to the block.

The fluid is incompressible and Newtonian, with density $\rho_f=1.18\times10^{-3}\,\mathrm{g/cm^3}$ and dynamic viscosity $\mu_f=1.82\times10^{-4}\,\mathrm{g/(cm\,s)}$. A uniform streamwise velocity $U=51.3\,\mathrm{cm/s}$ is prescribed at the inlet. The upper and lower boundaries are impermeable slip boundaries, with zero normal velocity and zero tangential traction. No-slip conditions are imposed on the fixed block, and fluid velocity matches solid velocity at the moving fluid--solid interface. The outlet has a zero-traction condition augmented by backflow stabilization with coefficient $\beta=0.2$, active where the flow reverses. Based on the specified parameters and block size, the Reynolds number is $\mathrm{Re}=\rho_f U D/\mu_f\simeq333$.

The beam is modeled as a neo-Hookean solid with density $\rho_s=0.1\,\mathrm{g/cm^3}$, Young's modulus $E=2.5\times10^6\,\mathrm{g/(cm\,s^2)}$, and Poisson's ratio $\nu=0.35$. These values correspond to a shear modulus $\mu_s=E/[2(1+\nu)]\simeq9.26\times10^5\,\mathrm{g/(cm\,s^2)}$, a first Lam\'e parameter $\lambda_s=E\nu/[(1+\nu)(1-2\nu)]\simeq2.16\times10^6\,\mathrm{g/(cm\,s^2)}$, and a solid-to-fluid density ratio of approximately $85$.

\begin{figure}[!htbp]
  \centering
  \includegraphics[width=\linewidth]{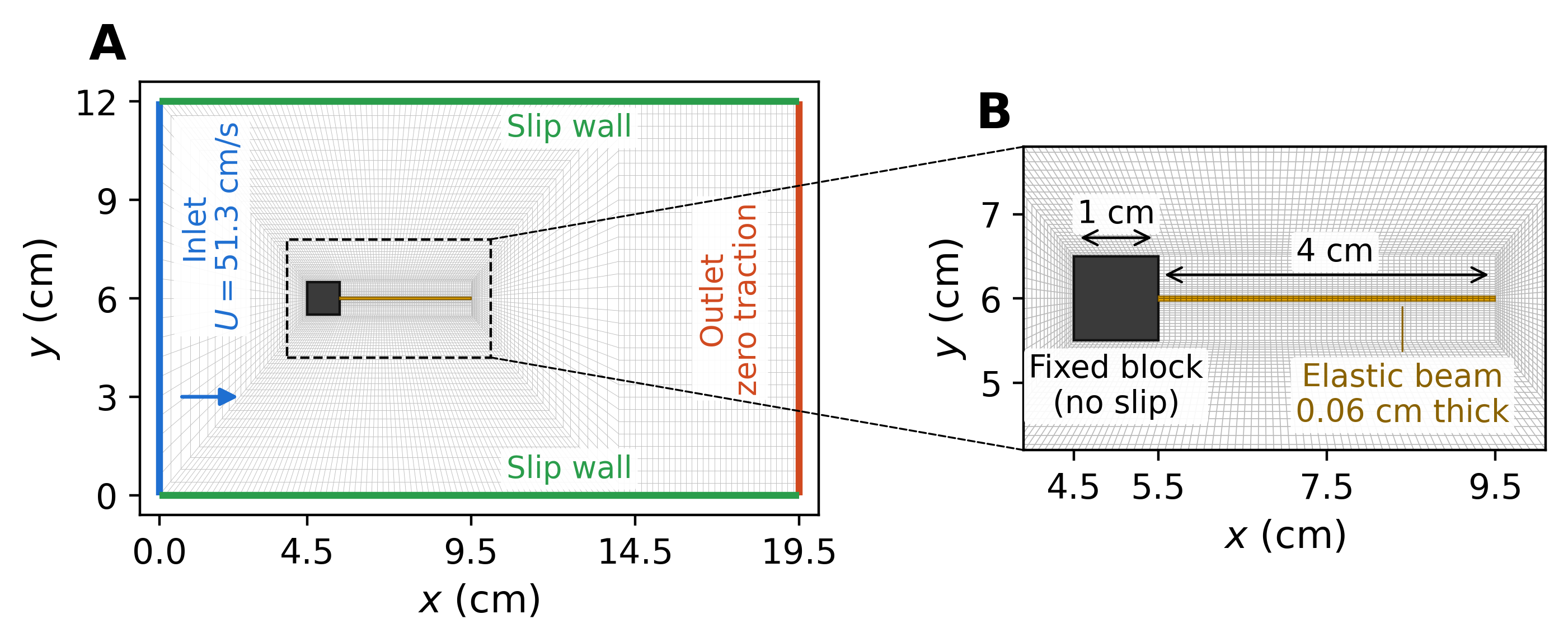}
  \caption{\textbf{Computational domain for the FSI benchmark.} (A) Fluid channel of $19.5\,\mathrm{cm}\times12\,\mathrm{cm}$ with its boundary conditions, shown with the fluid mesh. (B) Close-up of the fixed square block (side $1\,\mathrm{cm}$, centered at $(5,6)\,\mathrm{cm}$) and the elastic beam ($4\,\mathrm{cm}\times0.06\,\mathrm{cm}$, shaded) clamped to its rear face, with the solid mesh of the beam.}
  \label{fig:fsi-domain}
\end{figure}

\subsection*{Discretization and solution procedure}
The present simulation uses the same biquadratic quadrilateral mesh as the study of Esmaily-Moghadam et al.~\cite{moghadam2015}, with approximately $13{,}000$ elements across the fluid and solid domains. The mesh is refined near the block and beam and becomes coarser toward the outer boundaries. Fluid and solid nodes coincide at the interface, providing a conforming discretization.

The incompressible fluid equations are solved in an arbitrary Lagrangian--Eulerian formulation. The fluid and solid equations are coupled using the quasi-direct strategy, with the mesh-motion problem solved separately from the coupled fluid--solid system. Linear-elastic mesh motion with Jacobian-based stiffening accommodates beam deformation without remeshing. Time integration uses the generalized-$\alpha$ method with spectral radius $\rho_\infty=0.5$ and time step $\Delta t=10^{-3}\,\mathrm{s}$. The simulation is advanced for $10\,\mathrm{s}$ of physical time with svMultiPhysics v1.0.0.2 on 40 MPI processes. The prescribed relative nonlinear residual tolerance is $10^{-6}$, with $3$ to $10$ Newton iterations per time step; it was met at every time step, with an average of $4.8$ Newton iterations. The coupled fluid--solid linear systems are solved with GMRES preconditioned by the Trilinos ML algebraic multigrid, with a relative tolerance of $10^{-6}$ and at most $100$ iterations per solve. In practice most linear solves end at this iteration limit, so that Newton's method operates as an inexact Newton method. The mesh-motion equation is solved with the conjugate gradient method. The beam oscillation is established after approximately $4\,\mathrm{s}$. From then on, each second of simulated time requires $1377$--$1390\,\mathrm{s}$ of wall-clock time on 40 processes, with $5.0$ Newton iterations per time step, compared with $1114\,\mathrm{s}$ and $4.1$ Newton iterations per time step during the first second; for this FSI problem, the cost of the developed periodic response is approximately $25\%$ higher than that of the initial transient.

\clearpage
\subsection*{Comparison with published results}
After the initial transient, vortex shedding produces sustained transverse beam oscillations. Because the onset of this motion is sensitive to numerical perturbations~\cite{bazilevs2008fsi}, the comparison focuses on the established oscillatory response. Fig~\ref{fig:fsi-benchmark-tip} compares the transverse displacement of the beam tip over approximately three oscillation cycles. The histories are shifted in time to align their phases; the plotted time therefore denotes the comparison window rather than elapsed time from initialization.

\begin{figure}[!htbp]
  \centering

  \includegraphics[width=\linewidth]{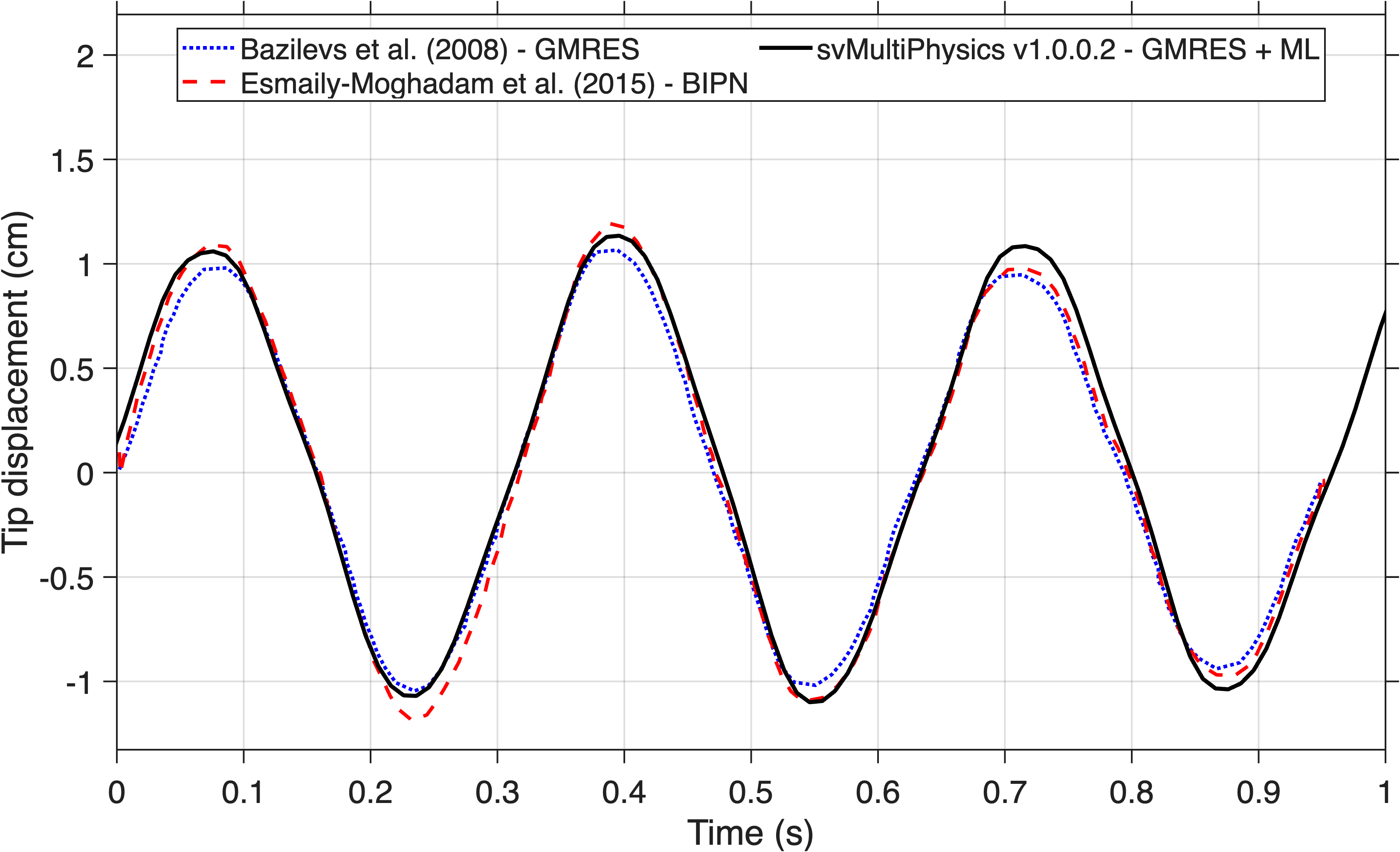}

  \caption{\textbf{Transverse beam-tip displacement for the FSI benchmark.}
  Solid black line: present svMultiPhysics v1.0.0.2 simulation (GMRES with ML preconditioning), with $\rho_\infty=0.5$.
  Dashed red line: published BIPN result of Esmaily-Moghadam et al.~\cite{moghadam2015} (2015), obtained with the predecessor Fortran implementation.
  Dotted blue line: isogeometric result of Bazilevs et al.~\cite{bazilevs2008fsi} (2008), using GMRES.
  The initial transient is excluded and the histories are shifted in time to align their phases.}
  \label{fig:fsi-benchmark-tip}
\end{figure}

The svMultiPhysics oscillation frequency was computed from the last eight complete oscillation periods, between $7.35$ and $9.94\,\mathrm{s}$, using the upward crossings of the mean beam-tip displacement. The reference frequencies were estimated by fitting the final available portions of the published displacement histories: $3.5$--$4.5\,\mathrm{s}$ for svFSI, and $0.0019815$--$0.94928\,\mathrm{s}$ for the Bazilevs et al. dataset, whose available window spans approximately $0.9473\,\mathrm{s}$. These intervals refer to the respective datasets' time coordinates. With $\rho_\infty=0.5$, svMultiPhysics yields a frequency of $3.0975\,\mathrm{Hz}$ and a period of $0.32284\,\mathrm{s}$, compared with $3.1179\,\mathrm{Hz}$ and $0.32073\,\mathrm{s}$ for svFSI (Fortran predecessor) at the same $\rho_\infty$, and $3.1584\,\mathrm{Hz}$ and $0.31662\,\mathrm{s}$ for the isogeometric result of Bazilevs et al. Relative to the respective reference frequencies, the svMultiPhysics frequency is lower by $0.65\%$ than svFSI and by $1.93\%$ than the Bazilevs et al. result. It is also within $0.9\%$ of the approximate period of $0.32\,\mathrm{s}$ (frequency $3.125\,\mathrm{Hz}$) reported by Esmaily-Moghadam et al.~\cite{moghadam2015}.

The close agreement in oscillation frequency, together with the similar displacement waveforms in Fig~\ref{fig:fsi-benchmark-tip}, supports consistency of the present FSI implementation with the predecessor solver and the published isogeometric benchmark.

\clearpage
\refstepcounter{benchmarksection}\label{app:ep-benchmark}
\section*{Appendix \thebenchmarksection. Electrophysiology Benchmark}

\subsection*{Problem definition}
We assessed the electrophysiology implementation using the benchmark proposed by Niederer et al. \cite{Niederer2011}. 
The benchmark considers a $20 \times 7 \times 3\ \mathrm{mm}^3$ slab of cardiac tissue~(Fig~\ref{fig:ep-benchmark-domain}) governed by the monodomain model and the ten Tusscher--Panfilov model of human epicardial myocytes. 
The fibers are aligned with the 20-mm axis, and all boundaries have zero normal flux. 
The tissue is transversely isotropic, with intracellular conductivities of $0.17$ and $0.019\ \mathrm{S\,m^{-1}}$ in the longitudinal and transverse directions, respectively, and corresponding extracellular conductivities of $0.62$ and $0.24\ \mathrm{S\,m^{-1}}$. 
Using a membrane surface-area-to-volume ratio of $140\ \mathrm{mm^{-1}}$ and a specific membrane capacitance of $0.01\ \mu\mathrm{F\,mm^{-2}}$, these values were converted to an isotropic diffusion $D_{\mathrm{iso}}=0.012571$ and $D_{\mathrm{f}}=0.082715 \mathrm{mm^2\,ms^{-1}}$ ($D_{\mathrm{s}}=D_{\mathrm{n}}=0$).
Cell-model parameters and initial conditions followed the benchmark specification. 
A $1.5 \times 1.5 \times 1.5\ \mathrm{mm}^3$ cube at the origin was stimulated for $2\ \mathrm{ms}$ with a volumetric current of $50000\ \mu\mathrm{A\,cm^{-3}}$.

\begin{figure}[!htbp]
  \centering
  \includegraphics[width=\linewidth]{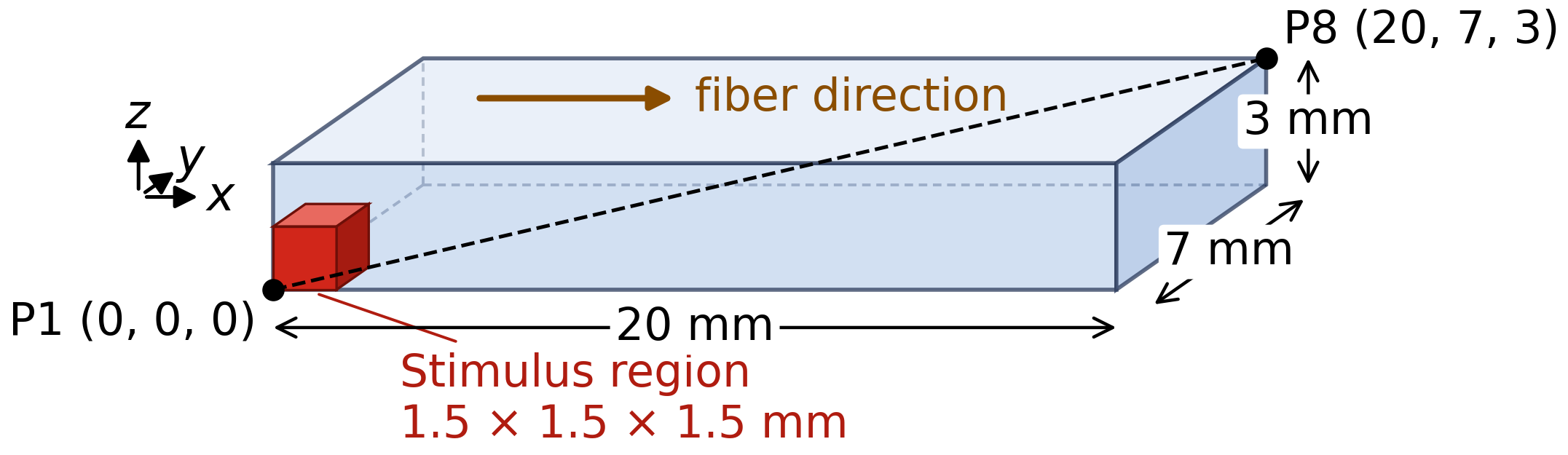}
  \caption{\textbf{Domain of the electrophysiology benchmark.} Slab of $20\times7\times3\ \mathrm{mm}^3$ with fibers along $x$. The stimulus is applied in the $1.5\times1.5\times1.5\ \mathrm{mm}^3$ cube at the corner P1 $=(0,0,0)$ (red); activation times are reported along the diagonal from P1 to the opposite corner P8 $=(20,7,3)\ \mathrm{mm}$ (dashed).}
  \label{fig:ep-benchmark-domain}
\end{figure}

\subsection*{Discretization and solution procedure}
For the results reported here, we discretized the slab with linear tetrahedral elements having target mean edge lengths $\Delta x$ of $0.5$, $0.2$, and $0.1\ \mathrm{mm}$. Each mesh was simulated using time steps $\Delta t$ of $0.05$, $0.01$, and $0.005\ \mathrm{ms}$, giving nine space--time discretizations. The linear system was solved using the conjugate-gradient method with a tolerance of $10^{-12}$. The ODE model was solved explicitly using a forward Euler approach. Given the operator splitting strategy, a single non-linear iteration was needed to reach tolerances below $10^{-12}$. Activation time was defined as the first upward crossing of $0\ \mathrm{mV}$ and was linearly interpolated between solution fields saved every $0.05\ \mathrm{ms}$.  All simulations were performed with svMultiPhysics v1.0.0.2.

\subsection*{Comparison with published results}
Fig~\ref{fig:ep-benchmark} presents the same three plots as Figures 2--4 of the original benchmark paper. For the finest space--time discretization ($\Delta x=0.1\ \mathrm{mm}$ and $\Delta t=0.005\ \mathrm{ms}$), the activation time at the point P8 $ =(20,7,3)$ was $40.25\ \mathrm{ms}$, within the $37.8$--$48.7\ \mathrm{ms}$ range reported across the original solvers at this resolution. At $\Delta t=0.005\ \mathrm{ms}$, the P8 activation times ranged from $38.42$ to $40.25\ \mathrm{ms}$ across the three mesh resolutions, indicating relatively modest sensitivity to spatial discretization, as observed for solvers A and H in the original study. The trends are also qualitatively consistent with more recent implementations in TorchCor \cite{zhou2026} (see Fig~3 in that paper) and openCARP \cite{opencarp} (see the tutorial in \cite{linder2026tutorial}).
These results support the ability of svMultiPhysics to reproduce the expected electrophysiological behavior in this benchmark.

\begin{figure}[!htbp]
  \centering

  \includegraphics[width=\linewidth]{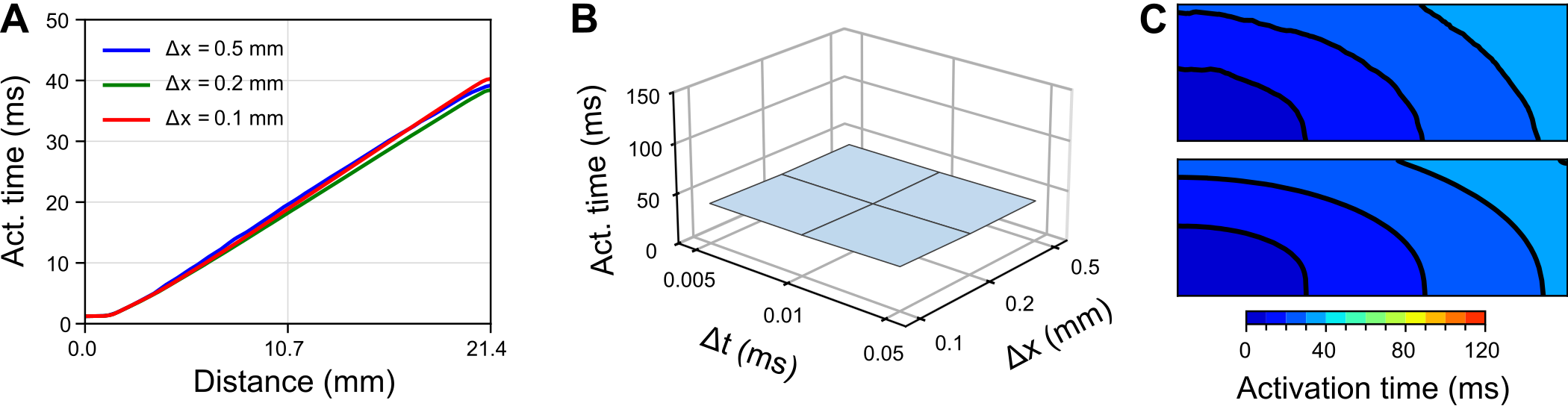}

  \caption{\textbf{Electrophysiology benchmark results.}
  (A) Activation time along the line from the origin (P1) to the farthest corner (P8) for the three spatial resolutions at $\Delta t=0.005\ \mathrm{ms}$. (B) Activation time at P8 for all combinations of $\Delta x$ and $\Delta t$. (C) Activation time maps on the benchmark plane (normal vector $(10.5,30,-140)$) for $\Delta x=0.5\ \mathrm{mm}$ (top) and $\Delta x=0.1\ \mathrm{mm}$ (bottom), both at $\Delta t=0.005\ \mathrm{ms}$.}
  \label{fig:ep-benchmark}
\end{figure}

\clearpage

\refstepcounter{benchmarksection}\label{app:cardiac-mechanics-benchmark}
\section*{Appendix \thebenchmarksection. Cardiac Mechanics Benchmark\\with Passive and Active Stress}

\subsection*{Problem definition}
For assessing the cardiac mechanics implementation in svMultiPhysics, we consider the benchmark proposed by Ar\'ostica et al. \cite{arostica2025benchmark}. 
In particular, we focus on "benchmark 1, Step 1", which models an idealized left ventricle~(Fig~\ref{fig:struct-domain}) subjected to time-dependent endocardial pressure and active contraction over one cardiac cycle.

Following the benchmark specifications, the myocardium was modeled as a hyperelastic, viscous, nearly incompressible material. 
The hyperelastic response was described by the Holzapfel--Ogden model with modified anisotropy: the isotropic term uses the isochoric first invariant, whereas the anisotropic terms use the full invariants $I_{4f}$, $I_{4s}$, and $I_{8fs}$~\cite{nolan2014}. 
The viscous term was represented using the potential-based viscous formulation of Chapelle et al.~\cite{chapelle2012}, and the volumetric response was represented using the Simo--Taylor formulation~\cite{simotaylor91}. 
The specific strain-energy density formulations can be found in the S1 Appendix.
All material parameters were set to the values specified in the original benchmark.

We used the active-stress model in Eq.~(11) of the main text, with active stress restricted to the fiber direction:
$\mathbf{S}_{\mathrm{act}} = T_{\mathrm{act}}(t)\,\mathbf{f}\otimes\mathbf{f}$.
Spring--dashpot Robin boundary conditions were applied at the base and epicardium using the parameters specified in the original benchmark.
The endocardial pressure $p(t)$ was prescribed using a Neumann boundary condition.
The endocardial pressure $p(t)$ and active tension ($T_{\mathrm{act}}(t)$) time traces were precomputed by solving the ODEs specified in the original benchmark. 
The myocardial fiber and sheet fields were generated following the procedure in the original paper.

\begin{figure}[!htbp]
  \centering
  \includegraphics[width=\linewidth]{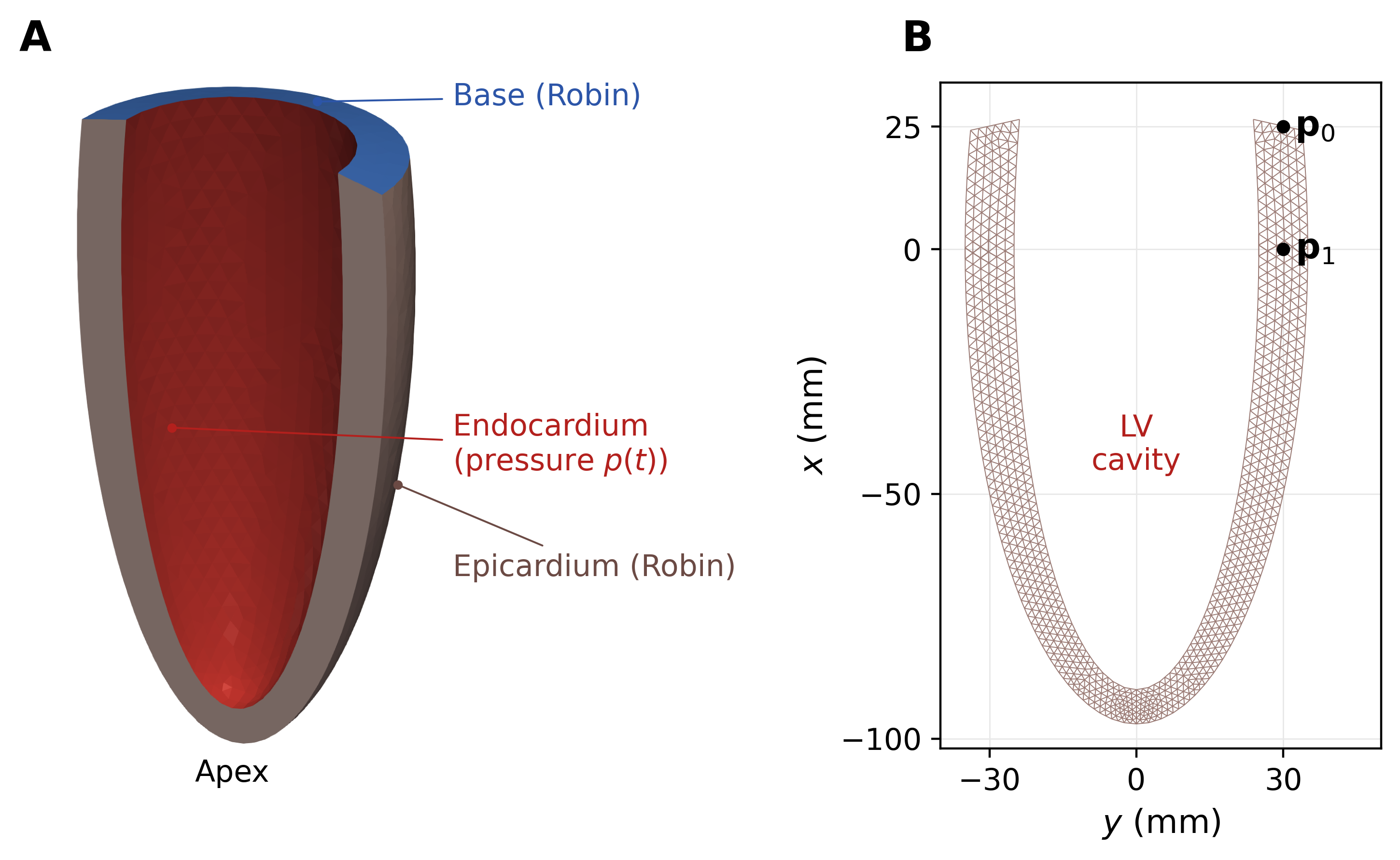}
  \caption{\textbf{Domain of the cardiac mechanics benchmark.} (A) Idealized left ventricle cut at $z=0$, showing the endocardium (red), where the pressure $p(t)$ is applied, and the epicardium (brown) and base (blue), where the Robin conditions are applied. (B) Section $z=0$ of the quadratic tetrahedral mesh, with the material points $\mathbf{p}_0$ and $\mathbf{p}_1$ at which the displacements are compared. Coordinates in mm.}
  \label{fig:struct-domain}
\end{figure}

\subsection*{Discretization and solution procedure}
We used the linear tetrahedral mesh directly from the benchmark database~\cite{arostica2024benchmarkdata}. A quadratic tetrahedral mesh was generated from it by adding nodes at the edge midpoints. 
We used a time step of $\Delta t=1\,\mathrm{ms}$ over the $1\,\mathrm{s}$ cycle. 
The linear systems were solved with GMRES using relative and absolute tolerances of $10^{-12}$. 
The relative nonlinear residual tolerance was set to $10^{-9}$.  The simulation was performed with svMultiPhysics v1.0.0.2.

\subsection*{Comparison with published results}
Fig.~\ref{fig:struct-benchmark} follows the layout of Fig.~10 in the original paper and shows the three displacement components at the material points $\mathbf{p}_0=(0.025,0.03,0)\,\mathrm{m}$ and $\mathbf{p}_1=(0,0.03,0)\,\mathrm{m}$. 
The gray band spans the minimum and maximum values reported by the solvers in the original benchmark at each time point. 
The svMultiPhysics solution is shown in red. 
The traces remain within the benchmark range for all displacement components at both points. 

These results demonstrate that svMultiPhysics can reproduce benchmark results for cardiac mechanics simulations involving complex material models and simulation conditions. Because the original benchmark included svFSI among the solvers, these findings also indicate consistency with the previous implementation.

\begin{figure}[!htbp]
  \centering

  \includegraphics[width=\linewidth]{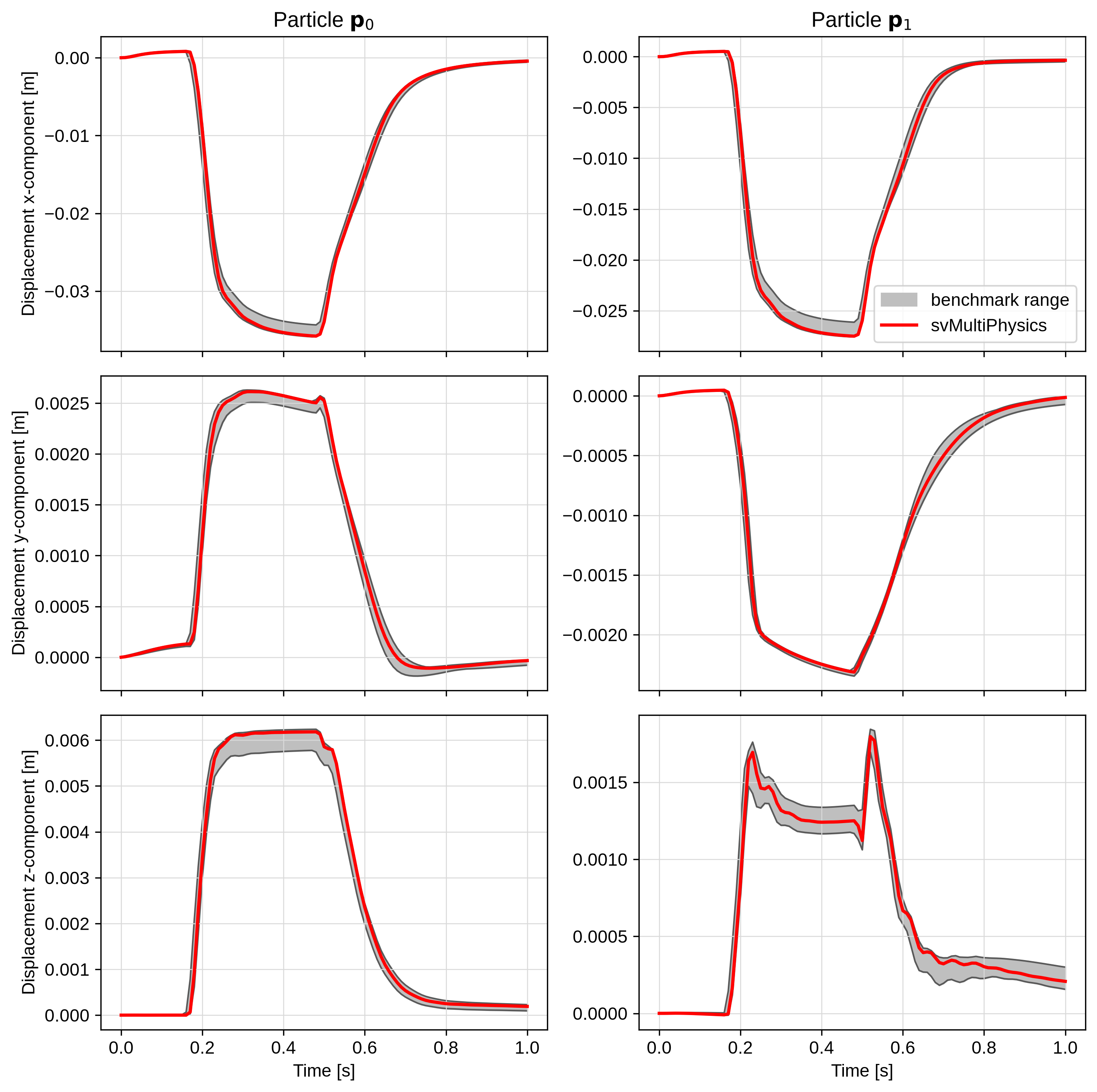}

  \caption{\textbf{Cardiac mechanics benchmark results.}
  Components of the displacement at material points $\mathbf{p}_0$ and $\mathbf{p}_1$. The gray band spans the range of the solutions reported in the original benchmark. The solution obtained with svMultiPhysics is shown in red.}
  \label{fig:struct-benchmark}
\end{figure}

\clearpage

\refstepcounter{benchmarksection}\label{app:whole-heart-coupling}
\section*{Appendix \thebenchmarksection. Whole-Heart FSI with Active Contraction}
\makeatletter
\setlength{\@fptop}{0pt}
\setlength{\@fpsep}{12pt}
\setlength{\@fpbot}{0pt plus 1fil}
\makeatother

We demonstrate the coupling of active myocardial contraction, intracardiac blood flow, valve function, and the systemic and pulmonary circulations using the whole-heart model described in Chapter~5 of Brown's dissertation~\cite{brown2026thesis}. The image-derived geometry includes all four cardiac chambers and the great vessels. The solid domain comprises the left and right atria, the biventricle, the great vessels, and an atrioventricular transition region. The formulation combines anisotropic passive tissue mechanics, region-specific myocardial fiber fields, and spatially varying epicardial spring--damper support. Fig~\ref{fig:whole-heart-setup} summarizes the coupled model.

Active contraction is driven by prescribed double-Hill stress waveforms, spatially uniform within each contractile region. In the biventricle, $70\%$ of the active stress acts along the fibers and $30\%$ along the sheet-normal direction. The fluid and solid equations are coupled monolithically in an arbitrary Lagrangian--Eulerian formulation: myocardial contraction drives wall motion and blood flow, while fluid tractions load the deforming tissue. The four valves are represented by resistive immersed surfaces whose opening and closing are triggered by local pressure and flow conditions. Their motion follows precomputed leaflet kinematics rather than an online solution of leaflet structural mechanics. A modular implicit interface couples the aortic, pulmonary arterial, pulmonary venous, and caval boundaries to a closed-loop lumped-parameter circulation solved by svZeroDSolver, exchanging flow rates and pressures.

The case was computed with a development version of svMultiPhysics, not the v1.0.0.2 release, and was advanced through five cardiac cycles of $0.69\,\mathrm{s}$ each, using $6900$ time steps of $5\times10^{-4}\,\mathrm{s}$. Representative snapshots illustrate ventricular filling and ejection through the moving valve surfaces (Fig~\ref{fig:whole-heart-flow}). Synchronized active-stress, pressure, volume, and valve-flow histories over the final cycle, together with atrial and ventricular pressure--volume loops, provide complementary measures of the coupled response (Fig~\ref{fig:whole-heart-hemodynamics}).

Over the final cycle, the left ventricle has an end-diastolic volume of $113.0\,\mathrm{mL}$, a stroke volume of $49.3\,\mathrm{mL}$, an ejection fraction of $43.6\%$, and a peak pressure of $89.7\,\mathrm{mmHg}$. The right ventricle has a stroke volume of $48.6\,\mathrm{mL}$, an ejection fraction of $32.8\%$, and a peak pressure of $36.6\,\mathrm{mmHg}$. The two stroke volumes agree to within $2\%$, and the ventricular volumes at the end of the cycle differ from those at its start by less than $0.6\,\mathrm{mL}$, indicating a nearly periodic state. The backward-flow volume over the final cycle is at most $2.6\%$ of the forward volume (aortic valve), and tricuspid backflow is $1.0\%$. Aortic pressure ranges from $37$ to $65\,\mathrm{mmHg}$, below physiological values, consistent with the preliminary circulatory parameterization discussed by Brown~\cite{brown2026thesis}. This example demonstrates active-mechanics--FSI--circulation coupling within svMultiPhysics; electrical propagation is not solved in this case. It is a demonstration of integrated functionality rather than a validation against clinical measurements.

\begin{figure}[!htbp]
  \centering
  \includegraphics[width=0.7\linewidth]{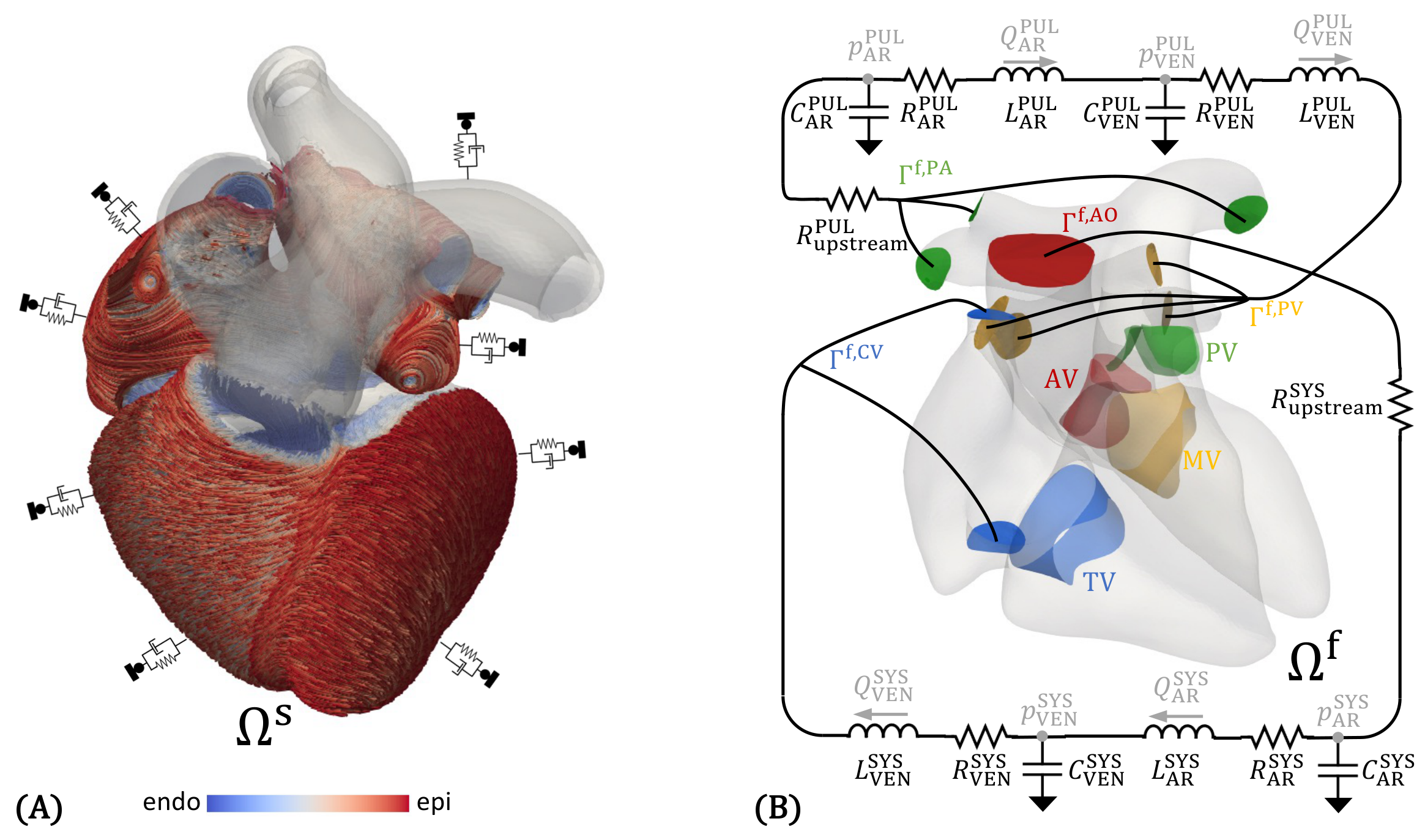}
  \caption{\textbf{Whole-heart model and circulatory coupling.}
  (A) Solid domain $\Omega^{\mathrm{s}}$ with rule-based myocardial fibers, colored from endocardium (blue) to epicardium (red), and spatially varying spring--damper (Robin) support on the epicardial and great-vessel surfaces. (B) Fluid domain $\Omega^{\mathrm{f}}$ with the mitral (MV), aortic (AV), tricuspid (TV), and pulmonary (PV) valves modeled as resistive immersed surfaces. The aortic ($\Gamma^{\mathrm{f,AO}}$), pulmonary arterial ($\Gamma^{\mathrm{f,PA}}$), pulmonary venous ($\Gamma^{\mathrm{f,PV}}$), and caval ($\Gamma^{\mathrm{f,CV}}$) boundaries are coupled to a closed-loop lumped-parameter model of the systemic (SYS) and pulmonary (PUL) circulations. Reproduced from Brown~\cite{brown2026thesis}, Fig.~5.2.}
  \label{fig:whole-heart-setup}
\end{figure}

\begin{figure}[!htbp]
  \centering
  \includegraphics[width=0.7\linewidth]{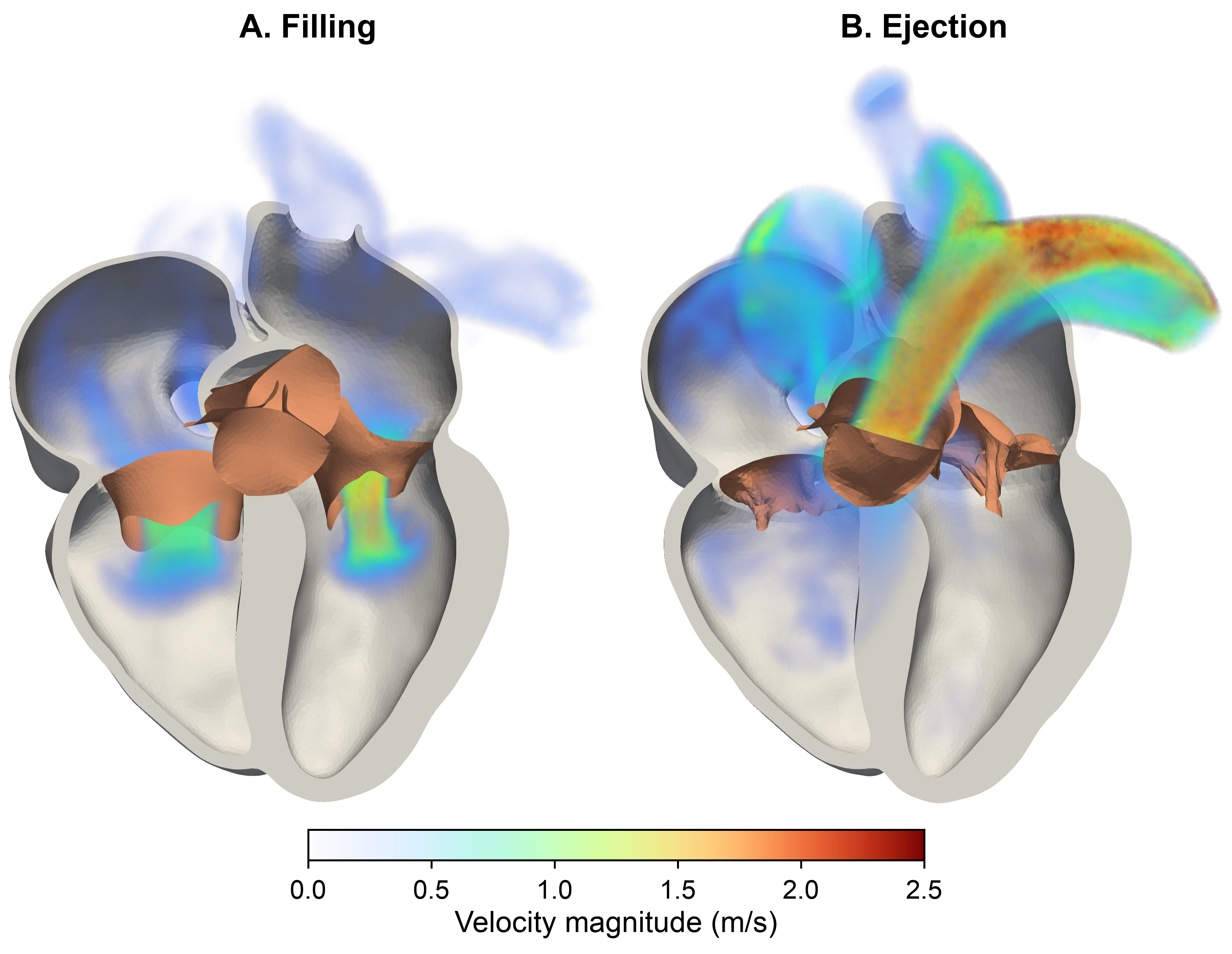}
  \caption{\textbf{Blood flow in the contracting whole-heart model.}
  Representative snapshots show (A) ventricular filling through the atrioventricular valves and (B) ejection into the great vessels. The cutaway wall is gray and immersed valve surfaces are orange. Blood is volume-rendered by velocity magnitude.}
  \label{fig:whole-heart-flow}

\end{figure}

\begin{figure}[!htbp]
  \centering
  \includegraphics[width=\linewidth,height=0.75\textheight,keepaspectratio]{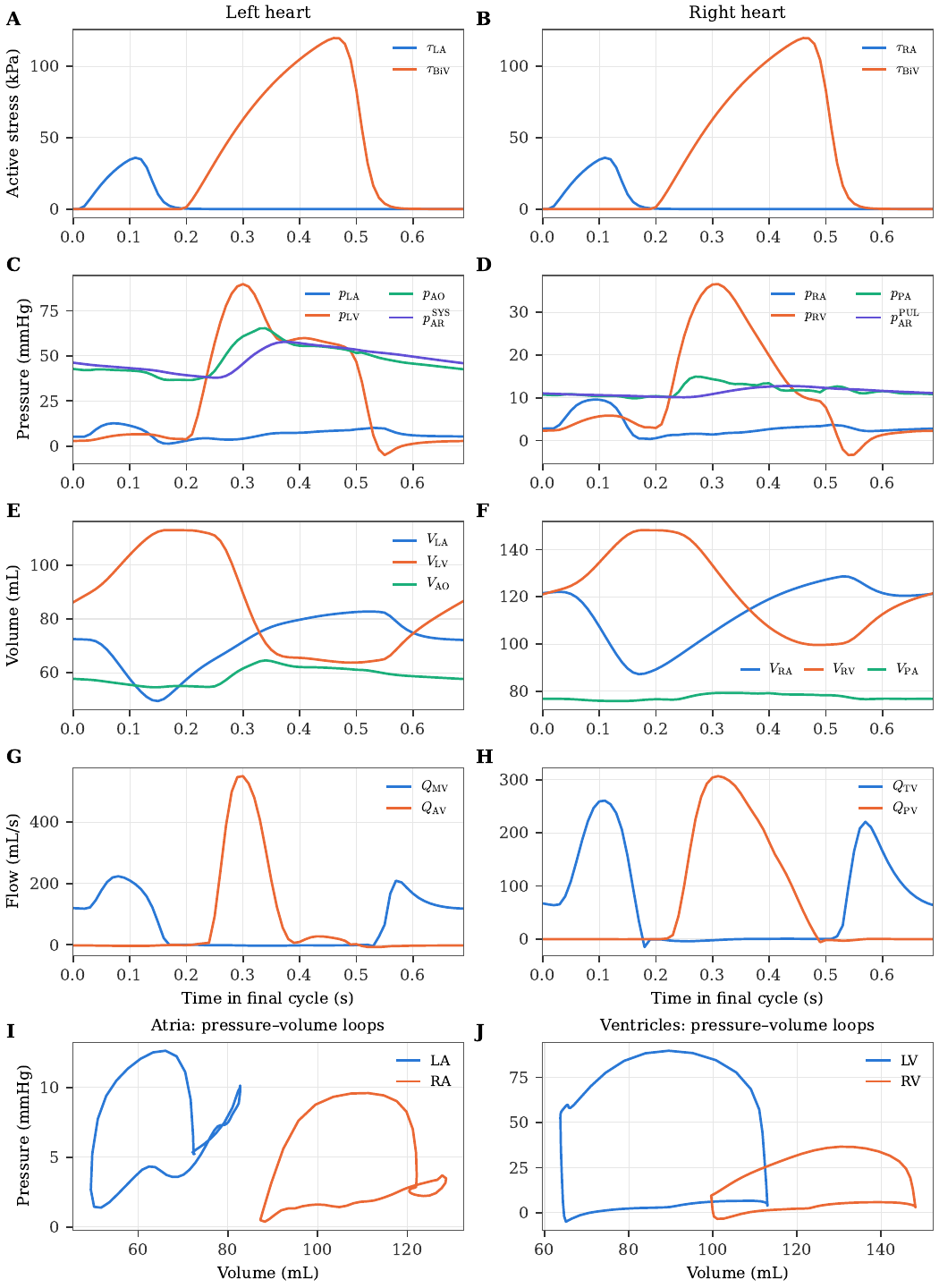}
  \caption{\textbf{Synchronized mechanical and circulatory response over the final cardiac cycle.}
  Left column, left heart; right column, right heart. (A, B) Prescribed active stress in the atria and biventricle. (C, D) Pressures in the atria, ventricles, and great vessels (averaged over the corresponding regions of $\Omega^{\mathrm{f}}$) and in the proximal systemic and pulmonary arterial compartments of the lumped-parameter model. (E, F) Chamber and great-vessel volumes. (G, H) Flow rates through the four valves. (I, J) Pressure--volume loops of the atria and ventricles. Time is measured from the start of the fifth cycle ($t=2.76\,\mathrm{s}$). The layout follows Brown~\cite{brown2026thesis}, Figs.~5.5--5.6.}
  \label{fig:whole-heart-hemodynamics}
\end{figure}

\clearpage

\refstepcounter{benchmarksection}\label{app:electromechanics}
\section*{Appendix \thebenchmarksection. Ventricular Electromechanics}

We demonstrate the electromechanical coupling described in the Solid mechanics section using an idealized left ventricle based on the geometry of the cardiac mechanics benchmark of Appendix~\ref{app:cardiac-mechanics-benchmark}~\cite{arostica2025benchmark}. Unlike Appendices~\ref{app:cardiac-mechanics-benchmark} and~\ref{app:whole-heart-coupling}, where the active tension is prescribed, here it is computed from the simulated intracellular calcium transient, so that electrical activation, active force generation, tissue deformation, and the circulation interact within a single simulation. This case demonstrates coupled functionality; it is not a benchmark comparison, because its activation sequence differs from the uniform activation of Appendix~\ref{app:cardiac-mechanics-benchmark}, and the model has not been calibrated to clinical data.

\subsection*{Problem definition}
Electrophysiology is described by the monodomain equation with the ten Tusscher--Panfilov ionic model (epicardial parameter set)~\cite{tentusscher2006} and anisotropic conductivity along the fiber and sheet directions. Activation is initiated by three endocardial stimuli near the apex, $120^\circ$ apart, each applied for $2\,\mathrm{ms}$. The myocardium is modeled with the modified-anisotropy Holzapfel--Ogden law~\cite{holzapfel2009,nolan2014}, a potential-based viscous term~\cite{chapelle2012}, and a quadratic volumetric penalty. The orifice of the prolate ellipsoid is closed by a soft plug at the top (Fig~\ref{fig:electromechanics}A) to ensure the pressure forces are properly balanced \cite{bucelli2025valves}. The active tension along the fibers is computed by the RDQ20-MF model~\cite{regazzoni2020}, with parameters from Fedele et al.~\cite{fedele2023}, adjusted to compensate for the lack of reference configuration recovery in this test case. Normal spring--damper Robin conditions are applied along the normal on the epicardium and on the basal surface. Springs are stiffer closer to the apex, to better reproduce the atrioventricular plane displacement \cite{strocchi2020}. The endocardium carries a follower pressure load determined by an open-loop lumped-parameter circulation solved by svZeroDSolver and coupled implicitly: a prescribed time-dependent atrial pressure, mitral and aortic valves with smooth (hyperbolic-tangent) pressure--flow laws, a proximal aortic compartment, and a three-element Windkessel outflow \cite{Brown2024ModularCoupling}.

\subsection*{Discretization and solution procedure}
The ventricle is discretized with 739102 linear tetrahedral elements and 144304 nodes, with average edge length 1.3 mm. The mechanics problem is advanced with a time step $1\,\mathrm{ms}$ and the ionic model is integrated with a fourth-order Runge--Kutta scheme and a substep of $0.1\,\mathrm{ms}$, for 3 heartbeats of duration $1000\,\mathrm{ms}$. The nonlinear mechanics problem is solved with Newton's method augmented by a backtracking line search, which prevents the stagnation that otherwise occurs when the valves switch state. The input files are provided in the Zenodo record listed in the Data availability section.

\subsection*{Results}

Simulation results are shown in Fig~\ref{fig:electromechanics}, representing propagation of the excitation front (Fig ~\ref{fig:electromechanics}B), active tension (Fig~\ref{fig:electromechanics}C), and traces of ventricular pressure and volume (Fig~\ref{fig:electromechanics}D). The simulated ventricle gets fully activated within 120 ms of the stimulus onset. The model is able to replicate successfully all heartbeat phases, thanks to the presence of valve elements in the lumped parameter model. The end-diastolic and end-systolic volumes are 147.5 ml and 82.6 ml, leading to a stroke volume of 64.9 ml and a 44\% ejection fraction, while peak left ventricular pressure is 131.7 mmHg. These results generally agree with the behavior of normal human hearts, with departures from normal ranges due to the model being idealized rather than calibrated on patient-specific data. Oveall, the simulation demonstrates that svMultiPhysics can effectively solve the coupled electrophysiology-mechanics-circulation models.

\begin{figure}[!htbp]
  \centering
  
  \includegraphics[width=\linewidth]{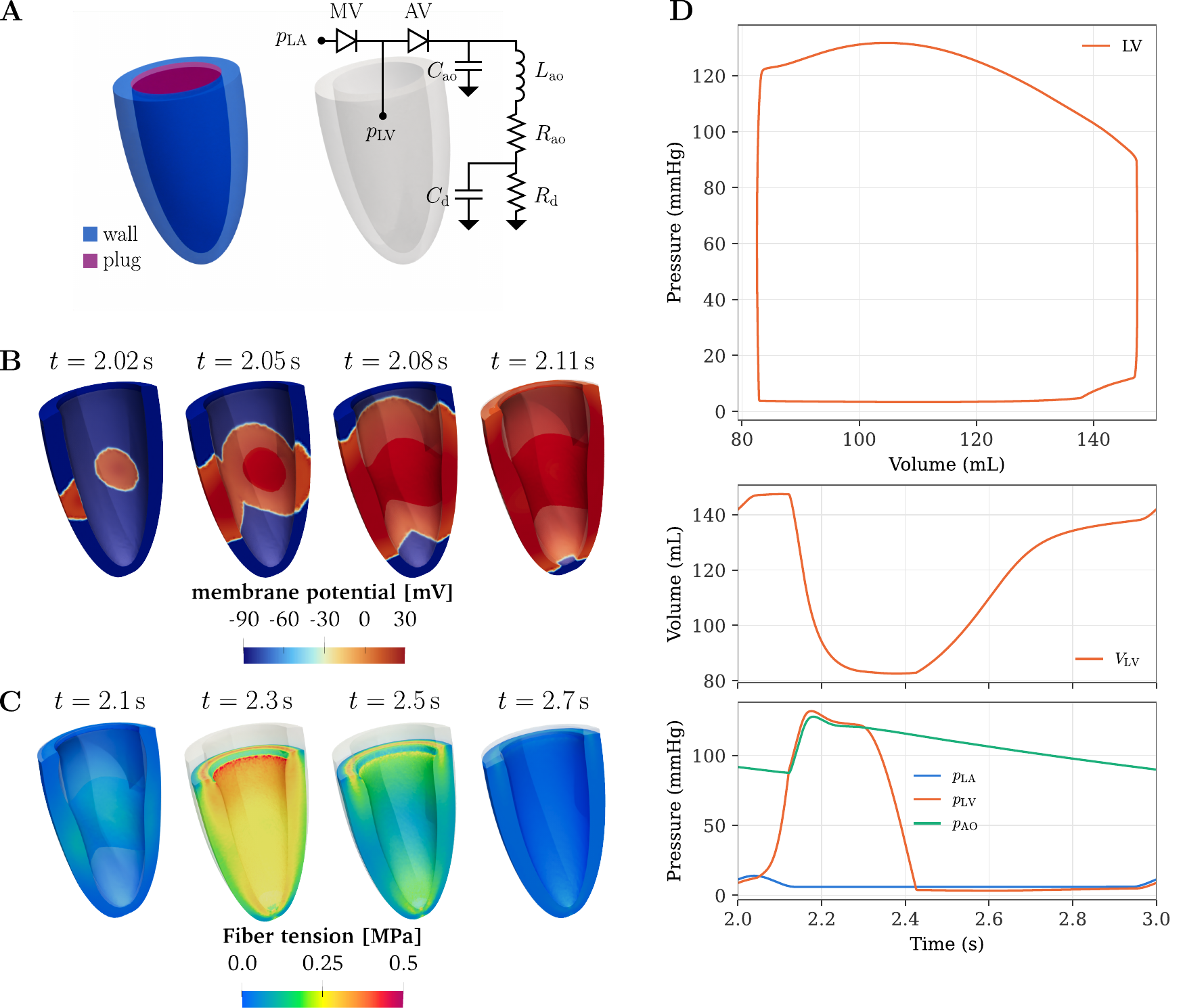}
  
  \caption{\textbf{Electromechanics of an idealized left ventricle.}
  The panels report results from the third simulated heartbeat. (A) Setup of the computational domain and of the lumped-parameters circulation model. The plug is removed from other visualization. (B) Snapshots of the transmembrane potential solution. (C) Snapshots of the active tension along fibers. The domain is warped by displacement, and the configuration at the beginning of the third beat is overlaid in transparency for comparison. (D) Left ventricular pressure-volume loop, ventricular volume and pressure over time.}
  \label{fig:electromechanics}
\end{figure}

\par
\endgroup

\section*{Supporting information}


\paragraph*{S1 Appendix}
\label{S1_Appendix} Constitutive material models implemented in svMultiPhysics.

\paragraph*{S2 website}
\label{svgithub}
{\bf https://github.com/SimVascular/svMultiPhysics} The GitHub repository for svMultiPhysics.

\paragraph*{S3 website}
\label{sv0d}
{\bf https://github.com/SimVascular/svZeroDSolver} The GitHub repository for svZeroDSolver.

\paragraph*{S4 website}
\label{svdockerhub}
{\bf https://hub.docker.com/r/simvascular/svmultiphysics/tags} The SimVascular DockerHub page contains images of pre--built svMultiPhysics and all necessary libraries.

\paragraph*{S5 website}
\label{svdoc}
{\bf https://simvascular.github.io} The SimVascular website contains an extensive documentation on the svMultiPhysics solver.

\paragraph*{S1 Table}\label{S1_Table}
{\bf Twenty-step AAA performance data.} Elapsed times, convergence counts, and Newton iterations for all 98 CPU runs with svMultiPhysics v1.0.0.2, together with the stage-level timing breakdown obtained with a local profiling build.

\paragraph*{S1 Fig}\label{S1_Fig}
{\bf Nonlinear and linear convergence histories of the AAA simulations.} Relative nonlinear residual and number of linear iterations per Newton iteration over the 20 time steps on 64 MPI processes (svMultiPhysics v1.0.0.2). (A) CFD with Trilinos GMRES + ML and FSILS BIPN. (B) FSI (fluid--solid equation) with Trilinos GMRES + ML and FSILS GMRES + diagonal preconditioning.

\paragraph*{S2 Fig}\label{S2_Fig}
{\bf AAA rigid-wall CFD over five cardiac cycles.} (A) Face-averaged pressure at the inlet and the two outlets; dashed lines show the cycle-averaged pressure at outlet 1. (B) Outlet 1 pressure and (C) outlet 1 flow rate, with the five cycles overlaid. (D) Time-stepping wall-clock time per cardiac cycle and for the first 20 time steps of each cycle, labeled with the mean number of Newton iterations per time step and the total number of Newton iterations, respectively.

\paragraph*{S1 Data}\label{S1_Data}
{\bf Performance data of the AAA and electrophysiology simulations.} Comma-separated values for all 98 AAA CPU runs and all 63 biventricular electrophysiology runs: elapsed times, converged steps, linear-solver cap counts, Newton and linear iteration totals (v1.0.0.2), and, for the AAA runs, stage timings of the profiling build. For the electrophysiology runs, the column \texttt{ep\_histor\_time\_step500\_s} gives the wall-clock time after 500 time steps shown in Figs~\ref{fig:scal-ep} and~\ref{fig:wall-ep}.

\section*{Data availability}
The svMultiPhysics source code is openly available under the BSD 3-Clause license at \url{https://github.com/SimVascular/svMultiPhysics}. The results in this paper were computed with the svMultiPhysics release v1.0.0.2 (\url{https://github.com/SimVascular/svMultiPhysics/releases/tag/v1.0.0.2}), except where stated otherwise in the text; the release provides the source code and prebuilt Ubuntu packages. The inputs of the simulations in this paper (meshes, fiber fields, boundary-condition data, and XML files of the AAA, biventricular electrophysiology, and benchmark cases from the appendix) are available on Zenodo at \url{https://doi.org/10.5281/zenodo.24486617} (the record will be made public after the article is accepted); the AAA geometry is derived from case 0031\_H\_ABAO\_AAA of the \href{https://www.vascularmodel.com/share.html?MzBOWTI4NU4_}{Vascular Model Repository}. These and the other benchmark cases are being added to the svMultiPhysics repository together with a benchmark harness (\href{https://github.com/SimVascular/svMultiPhysics/pull/736}{pull request \#736}; \href{https://github.com/SimVascular/svMultiPhysics/issues/662}{issue \#662}). The data underlying the performance figures are provided in \nameref{S1_Data}. The patient-specific whole-heart model of Appendix~\ref{app:whole-heart-coupling} is not publicly available; its inputs are available from the corresponding author on reasonable request.
Docker images with pre-built svMultiPhysics and its dependencies are available at \url{https://hub.docker.com/r/simvascular/svmultiphysics/tags}.

\section*{Acknowledgments}
This work was supported by the National Science Foundation under
award number 2310909. We acknowledge the computational resources and technical support provided by the Stanford Research Computing Center and the Texas Advanced Computing Center, which enabled the simulations and performance studies presented in this work.

\nolinenumbers

\end{document}